\documentclass{article}

\pdfoutput=1
\usepackage{arxiv}

\usepackage[utf8]{inputenc} 
\usepackage[T1]{fontenc}    
\usepackage[hidelinks]{hyperref}       
\usepackage{url}            
\usepackage{booktabs}       
\usepackage{amsfonts}       
\usepackage{nicefrac}       
\usepackage{microtype}      
\usepackage{lipsum}
\usepackage{graphicx}
\graphicspath{ {./images/} }
\usepackage{natbib}
\usepackage{amsmath}
\usepackage{makecell}
\usepackage{adjustbox}
\usepackage[table]{xcolor}
\usepackage{enumitem}

\usepackage{multirow}
\usepackage[title]{appendix}

\def\bSig\mathbf{\Sigma}

\title{Estimating SARS-CoV-2 Seroprevalence}

\author{
	Samuel P. Rosin \\ 
	Department of Biostatistics \\ University of North Carolina at Chapel Hill \\ 
	Chapel Hill, NC \\
	\texttt{srosin@live.unc.edu} \\
	\And 
	Bonnie E. Shook-Sa \\
	Department of Biostatistics \\ University of North Carolina at Chapel Hill \\ 
	Chapel Hill, NC \\
	\And 
	Stephen R. Cole \\
	Department of Epidemiology \\ University of North Carolina at Chapel Hill \\ 
	Chapel Hill, NC \\
	\And
	Michael G. Hudgens \\
	Department of Biostatistics \\ University of North Carolina at Chapel Hill \\ 
	Chapel Hill, NC 
}

\renewcommand{\headeright}{}
\renewcommand{\undertitle}{}
\renewcommand{\shorttitle}{Estimating SARS-CoV-2 Seroprevalence}

\hypersetup{
	colorlinks,
linkcolor={red!50!black},
citecolor={blue!50!black},
urlcolor={blue!80!black},
	pdftitle={Estimating SARS-CoV-2 Seroprevalence},
	pdfsubject={stat.AP, stat.ME},
	pdfauthor={Samuel P.~Rosin, Bonnie E.~Shook-Sa, Stephen R.~Cole, Michael G.~Hudgens},
	pdfkeywords={Covid-19, diagnostic tests, estimating equations, seroepidemiologic studies, standardization},
}

\begin{document}
	
	\maketitle
	
	\begin{abstract}{
				Governments and public health authorities use seroprevalence studies to guide responses to the COVID-19 pandemic. Seroprevalence surveys estimate the proportion of individuals who have detectable SARS-CoV-2 antibodies. However, serologic assays are prone to misclassification error, and non-probability sampling may induce selection bias. In this paper, nonparametric and parametric seroprevalence estimators are considered that address both challenges by leveraging validation data and assuming equal probabilities of sample inclusion within covariate-defined strata. Both estimators are shown to be consistent and asymptotically normal, and consistent variance estimators are derived. Simulation studies are presented comparing the estimators over a range of scenarios. The methods are used to estimate SARS-CoV-2 seroprevalence in New York City, Belgium, and North Carolina.}
	\end{abstract}
	
	
	\section{Introduction}
	\label{sec:intro}
	
	Estimating the proportion of people who have antibodies to SARS-CoV-2 is useful for tracking the pandemic's severity and informing public health decisions \citep{arora_serotracker_2021}. Individuals may have detectable antibodies for different reasons, including prior infection or vaccination. Antibody levels within a person are dynamic, typically increasing after an infection or vaccination, and then eventually decreasing (waning) over time. Thus individuals may not have detectable antibodies if never (or very recently) infected or vaccinated, or if their antibody levels have waned below the limit of detection of the assay being employed. To the extent that antibody levels are associated with protection from infection with SARS-CoV-2 or COVID-19 disease \citep{earle_evidence_2021, khoury_neutralizing_2021}, seroprevalence estimates may be helpful in modeling the fraction of a population which may be immune or less susceptible to COVID-19. Likewise, cross-sectional seroprevalence estimates, combined with certain modeling assumptions and other data, may permit inference about other parameters such as the cumulative incidence of previous SARS-CoV-2 infection, infection fatality rate, or attack rate \citep{takahashi_sars-cov-2_2021, shioda_estimating_2021, buss_three-quarters_2021, perez-saez_persistence_2021, brazeau_report_2020}.
	
	Unfortunately, seroprevalence studies often suffer from at least two sources of bias: measurement error due to false positives and negatives, and selection bias due to non-probability sampling designs. Typically, blood tests for antibodies result in a continuous measure of a particular antibody response, such as that of immunoglobulin G, M, or A (IgG, IgM, or IgA). Dichotomizing antibody responses using a cutoff value almost always produces misclassification bias in the form of false positives and false negatives \citep{bouman_estimating_2021}. The following example from Sempos and Tian (2021) demonstrates how this measurement error can lead to biased seroprevalence estimates. Suppose that true seroprevalence is 1\% and antibody tests are performed using an assay which perfectly identifies true positives as positive, so with 100\% sensitivity, and nearly perfectly identifies true negatives as negative, with 99\% specificity. Despite this assay’s high sensitivity and specificity, it is straightforward to show that naively using the sample proportion of positive test results as a seroprevalence estimator would, in expectation, lead to a seroprevalence estimate of nearly 2\% rather than 1\%. To account for measurement error, sensitivity and specificity can be estimated and incorporated into the seroprevalence estimator, e.g., using the method popularized by \citet{rogan_estimating_1978} (see also \citet{levy_three-population_1970, marchevsky_re_1979}).

	Many seroprevalence studies are conducted by non-probability sampling methods, which may lead to selection bias when characteristics that drive participation in the study are also risk factors for SARS-CoV-2 infection. Probability-based sampling studies are ideal because they are representative by design and often lead to less biased estimates than non-probability samples with post-hoc statistical adjustments \citep{shook-sa_estimation_2020, accorsi_how_2021}. However, probability-based sampling may not always be feasible due to time and cost constraints. For this reason, seroprevalence studies often utilize convenience sampling by, for example, drawing blood samples from routine clinic visitors \citep[e.g.,][]{barzin_sars-cov-2_2020, stadlbauer_repeated_2021} or using residual sera from blood donors \citep[e.g.,][]{uyoga_seroprevalence_2021} or commercial laboratories \citep[e.g.,][]{bajema_estimated_2021}. Convenience sample based estimators often assume that each person in a covariate-defined stratum has an equal probability of being in the sample \citep{elliott_inference_2017}. Under this assumption, population seroprevalence of SARS-CoV-2 can be estimated with direct standardization \citep{havers_seroprevalence_2020, barzin_sars-cov-2_2020, cai_exact_2022}, though weighting methods such as calibration can be used \citep[e.g.,][]{bajema_estimated_2021}.
	
	In this paper, methods are considered which combine standardization and the Rogan-Gladen adjustment to account for both measurement error and selection bias. The article is organized as follows. Section~\ref{sec:methods_rg} reviews prevalence estimation under measurement error. Nonparametric and parametric standardized prevalence estimators and their large-sample properties are described in Section~\ref{sec:selection_bias}. Section~\ref{sec:sims} presents simulation studies to evaluate the empirical bias and 95\% confidence interval (CI) coverage of the standardized estimators across a range of assay characteristics and bias scenarios. The methods are then applied in Section~\ref{sec:data_analysis} to three studies that estimate seroprevalence of SARS-CoV-2 in 2020 among all residents of New York City (NYC), all residents of Belgium, and asymptomatic residents of North Carolina.  Section~\ref{sec:discuss} concludes with a discussion. Proofs are in the Appendices.
	
	\section{Seroprevalence estimation under measurement error}
	\label{sec:methods_rg}
	
	\subsection{Problem setup}
	
	Let the true serology status for an individual in the target population be denoted by $Y$, with $Y=1$ if the individual has antibodies against SARS-CoV-2 and $Y=0$ otherwise. Our goal is to draw inference about the population seroprevalence $\pi=P(Y=1)$. Because of error in the serology assay, $Y$ is not observed directly. Let the result of the serology assay be denoted by $X$, with $X=1$ if the individual tests positive (according to the antibody assay used) and $X=0$ otherwise. Three key quantities are sensitivity, the probability that a true positive tests positive, denoted by $\sigma_e=P(X=1 \mid Y=1)$; specificity, the probability that a true negative tests negative, denoted by $\sigma_p = P(X=0 \mid Y=0)$; and the population expectation of the serology assay outcome, denoted by $\rho=\mathbb E(X)=P(X=1)$. Unless the assay has perfect sensitivity and specificity with $\sigma_e=\sigma_p=1$, $\rho$ typically will not equal $\pi$ and $X$ will be a misclassified version of $Y$. 
	
The sensitivity and specificity of a diagnostic test are commonly estimated by performing the assay on `validation' samples of known true positives and true negatives, respectively. Specifically, measurements are taken on $n_1$ independent and identically distributed (iid) units from strata of the population where $Y=1$ and on $n_2$ iid units from strata where $Y=0$.  Thus $n_1$ copies of $X$ are observed to estimate sensitivity and $n_2$ copies of $X$ are observed to estimate specificity. In the COVID-19 setting, samples from patients who had a case confirmed with reverse transcription polymerase chain reaction (PCR) testing are often assumed to be true positives. Remnant blood samples that were drawn in 2019 or earlier are often assumed to be true negatives. To estimate seroprevalence in a target population, a `main' study with $n_3$ iid copies of $X$ is then conducted, among which true infection status is unknown.
	
	Assume, as is realistic in many SARS-CoV-2 studies, that there is no overlap between the units in each of the three studies. Let $\delta_i$ be an indicator of which study the $i$th individual's sample $X_i$ is from, with $\delta_i=1$ for the sensitivity study, $\delta_i=2$ for the specificity study, and $\delta_i=3$ for the main study. Note that $\sum I(\delta_i=j)=n_j$ for $j=1,2,3$, where $n = n_1 + n_2 + n_3$ and here and throughout summations are taken from $i=1$ to $n$ unless otherwise specified. Assume $n_j/n \to c_j \in (0,1)$ as $n\to\infty$. 
	
	\subsection{Estimators and statistical properties}
	\label{sec:methods_rg_estimators}
	
	Let $\theta=(\sigma_e,\sigma_p,\rho,\pi)^T$. Consider the estimator $\hat \theta = (\hat \sigma_e, \hat \sigma_p,\hat \rho, \hat \pi_{RG})^T$, where $\hat \sigma_e = n_1^{-1}\sum I(\delta_i=1) X_i $, $\hat \sigma_p = n_2^{-1} \sum I(\delta_i=2) (1-X_i)$, $\hat \rho = 
	n_3^{-1}\sum I(\delta_i=3) X_i$, and $\hat \pi_{RG}=(\hat \rho + \hat \sigma_p -1) / (\hat \sigma_e + \hat \sigma_p -1)$. The prevalence estimator $\hat \pi_{RG}$ is motivated by rearranging the identity that $\rho = \pi \sigma_e + (1 - \pi)(1 - \sigma_p)$ and is sometimes referred to as the Rogan-Gladen (1978) estimator. Note the sample proportions $\hat \sigma_e$, $\hat \sigma_p$, and $\hat \rho$ are maximum likelihood estimators (MLEs) for $\sigma_e$, $\sigma_p$, and $\rho$, respectively, so $\hat \pi_{RG}$ is a function of the MLE of $(\sigma_e, \sigma_p, \rho)$ (see Appendix~\ref{sec:mle} for details).
	
	The estimator $\hat \theta$ can be expressed as the solution (for $\theta$) to the estimating equation vector \[ \sum \psi(X_i;\delta_i, \theta) = 
	\left( 
	\begin{array}{l}
		\sum \psi_e(X_i; \delta_i, \theta) \\
		\sum \psi_p(X_i; \delta_i, \theta) \\
		\sum \psi_{\rho}(X_i; \delta_i, \theta)\\
		\psi_{\pi}(X_i; \delta_i, \theta)
	\end{array}
	\right)
	= 
	\left( 
	\begin{array}{l}
		\sum I(\delta_i=1)(X_i - \sigma_e) \\
		\sum I(\delta_i=2)\{(1-X_i) - \sigma_p\} \\
		\sum I(\delta_i=3)(X_i - \rho) \\
		(\rho + \sigma_p - 1) - \pi(\sigma_e + \sigma_p -1)
	\end{array}
	\right) = 0 \] 
	where here and below 0 denotes a column vector of zeros. Since the samples were selected from three different populations, the data $X_1,\dots,X_n$ are not identically distributed and care must be taken to derive the large sample properties of $\hat \theta$. In Appendix~\ref{sec:proofs_rg}, the estimator $\hat \theta$ is shown to be consistent and asymptotically normal. Specifically, as $n \to \infty$, $\sqrt{n}(\hat \theta - \theta) \to_{d} \mathcal{N}\left(0, \mathbb A(\theta)^{-1}
	\mathbb B(\theta) 
	\mathbb A(\theta)^{-T}\right)$ and  $\sqrt{n}(\hat{\pi} - \pi) \to_{d} \mathcal{N}\left(0, V_{\pi,RG}\right)$ assuming $\sigma_e > 1 - \sigma_p$ (as discussed below),
	where $\mathbb A(\theta)^{-1}
	\mathbb B(\theta) 
	\mathbb A(\theta)^{-T}$ is a covariance matrix with bottom right element 
	\begin{equation}
		V_{\pi,RG} =
		\left\{
		\frac{\pi^2 \sigma_e (1-\sigma_e)}
		{c_1}
		+
		\frac{ (1-\pi)^2 \sigma_p (1-\sigma_p)}{c_2} 
		+ 
		\frac{\rho(1-{\rho})}{c_3}
		\right\} 
		(\sigma_e + \sigma_p -1)^{-2}.
		\label{eq:varpi}
	\end{equation}
	
	The proof of consistency and asymptotic normality is similar to proofs from standard estimating equation theory \citep[e.g.,][Equation 7.10]{boos_2013}, but because the data are not identically distributed the Lindeberg-Feller Central Limit Theorem (CLT) is used in place of the classical Lindeberg-L\'evy CLT. Note that the asymptotic variance \eqref{eq:varpi} consists of three components corresponding to the sensitivity, specificity, and main studies. In some circumstances, investigators may be able to decrease the variance of $\hat \pi_{RG}$ by increasing the sample sizes of the sensitivity or specificity studies compared to the main study \citep{larremore_jointly_2020}.
	
	Let $\hat V_{\pi, RG}$ denote the plug-in estimator defined by replacing $\sigma_e, \sigma_p, \rho, \pi$, and $c_j$ in \eqref{eq:varpi} with $\hat \sigma_e, \hat \sigma_p, \hat \rho, \hat \pi_{RG}$, and $n_j / n$ for $j = 1, 2, 3$, and note that $\hat V_{\pi, RG} / n$ is the variance estimator proposed by \citet{rogan_estimating_1978}. By the continuous mapping theorem, $\hat V_{\pi, RG}$ is consistent for the asymptotic variance assuming $\sigma_e > 1 - \sigma_p$ and can be used to construct Wald-type CIs that asymptotically attain nominal coverage probabilities. 
In finite samples, Wald-type CIs can sometimes have erratic coverage properties when estimating a single binomial parameter \citep{brown_interval_2001, dean_evaluating_2015}. In Section~\ref{sec:sims}, simulations are conducted to assess the performance of the Wald-type CIs in seroprevalence estimation scenarios. Alternative approaches for constructing CIs are discussed in Section~\ref{sec:discuss}.
	
	
	\subsection{Truncation into $[0,1]$}
	In finite samples $\hat \pi_{RG}$ sometimes yields estimates outside of $[0,1]$ when (i) $ \hat \sigma_e < 1 - \hat \sigma_p$, (ii) $\hat \rho < 1 - \hat \sigma_p$, or (iii) $\hat \rho > \hat \sigma_e$. Indeed, (ii) occurred in the ScreenNC study discussed in Section~\ref{sec:screennc}. Estimates are typically truncated to be inside $[0,1]$ because the true population prevalence must exist in $[0,1]$ \citep{hilden_further_1979}. In this article, all point estimates and bounds of interval estimates are so truncated. Note, though, that as the three sample sizes grow large the estimator $\hat \pi_{RG}$ yields estimates inside $[0, 1]$ almost surely unless $\sigma_e < 1 - \sigma_p$. In practice, settings where $\sigma_e < 1 - \sigma_p$ may be very unlikely; in such scenarios, the probability of a positive test result is higher for seronegative persons than for seropositive persons, so such a measurement instrument performs worse in expectation than random guessing. Throughout this manuscript, it is assumed that $\sigma_e > 1 - \sigma_p$.

	\section{Standardized seroprevalence estimation}
	\label{sec:selection_bias}
	
	\subsection{Problem setup}
	
	In some settings it may not be reasonable to assume the $n_3$ copies of $X$ from the main study constitute a random sample from the target population. Suppose instead that for each copy of $X$ a vector of discrete covariates $Z$ is observed, with $Z$ taking on $k$ possible values $z_1, \ldots, z_k$. The covariates $Z$ are of interest because seroprevalence may differ between the strata; for instance, $Z$ might include demographic variables such as age group, race, or gender. Denote the mean of $X$ in the $j$th stratum as $\rho_j = P(X=1\mid Z=z_j)$ and the sample size for the $j$th stratum as $n_{z_j}=\sum I(\delta_i=3,Z_i=z_j)$, so $\sum_{j=1}^kn_{z_j}=n_3$. 
	
	The distribution of strata in the target population, if known, can be used to standardize estimates so they are reflective of the target population (for a review of direct standardization, see \citet[][Chapter 15]{van_belle_biostatistics_2004}). Denote the proportion of the target population comprised by the $j$th stratum as $\gamma_j = P(Z=z_j)$ and suppose that these stratum proportions are known with each $\gamma_j>0$ and $\sum_{j=1}^k \gamma_j=1$. The stratum proportions are commonly treated as known based on census data or large probability-based surveys \citetext{\citealp[][Ch.~4.4]{lohr_sampling_2010}; \citealp[][Ch.~2.6]{korn_analysis_1999}. Alternatively, $\gamma_1, \dots, \gamma_k$ could be estimated, e.g., from a random sample of the target population, and the estimator of the seroprevalence estimator's variance could be appropriately adjusted to reflect the uncertainty in these estimated proportions.} 
	
	Assume that all persons in a covariate stratum defined by $Z$ have the same probability of inclusion in the sample. Then the covariates $Z$ in the main study sample have a multinomial distribution with $k$ categories, sample size $n_3$, and an unknown sampling probability vector $(s_1,\dots,s_k)^T$ where $\sum _{j=1}^k s_{j}=1$. For $j=1, \dots, k$, the probability $s_j$ indicates the chance of a sampled individual being in stratum $j$. Note that if the main study were a simple random sample from the target population, then the sampling probabilities would be equal to the stratum proportions (with $s_j=\gamma_j$ for $j=1,\dots,k$). 
	
	\subsection{Nonparametric standardization} 
	\label{sec:standardization} 
	First, consider a seroprevalence estimator which combines nonparametric standardization and the Rogan-Gladen adjustment to account for both selection bias and measurement error. Notice that $\rho$ is a weighted average of the stratum-conditional means $\rho_j$ where each weight is a known stratum proportion $\gamma_j$, i.e., $\rho=\sum_{j=1}^k \rho_j\gamma_j$. A nonparametric standardization estimator for $\rho$ using the sample stratum-conditional prevalences $\hat \rho_j = n_{z_j}^{-1}\sum I(Z_i=z_j,\delta_i=3)X_i$ for $j = 1, \dots, k$ is $\hat \rho_{SRG} = \sum_{j=1}^k \hat \rho_j \gamma_j$. A standardized prevalence estimator accounting for measurement error is $\hat \pi_{SRG} = (\hat \rho_{SRG}+\hat \sigma_p-1)/(\hat \sigma_e + \hat \sigma_p -1)$, which has been used in SARS-CoV-2 seroprevalence studies \citep{havers_seroprevalence_2020, barzin_sars-cov-2_2020, cai_exact_2022}.

	Let $\theta_s=(\sigma_e, \sigma_p, \rho_1,\dots,\rho_k,\rho,\pi)^T$. The estimator $\hat \theta_s=(\hat \sigma_e, \hat \sigma_p, \hat \rho_1,  \dots, \hat \rho_k, \hat \rho_{SRG}, \hat \pi_{SRG})^T$ solves the vector $\sum \psi(X_i,Z_i; \delta_i,\theta_s)=\left(\sum \psi_e, \sum \psi_p, \sum \boldsymbol\psi_{\boldsymbol\rho}, \psi_\rho,\psi_\pi\right)^T=0$ of estimating equations,
	where $\sum \psi_e, \sum \psi_p$, and $\psi_\pi$ are defined in Section~\ref{sec:methods_rg}; $\sum \boldsymbol\psi_{\boldsymbol\rho}$ is a $k$-vector with $j$th element $\sum \psi_{\rho_j}=\sum I(Z_i=z_j, \delta_i=3)(X_i-\rho_j)$; and $\psi_\rho=\sum_{j=1}^k\rho_j\gamma_j-\rho$. It follows that $\hat \theta_s$ is consistent and asymptotically normal and that $\sqrt{n}(\hat{\pi}_{SRG} - \pi) \to_{d} \mathcal{N}\left(0, V_{\pi,SRG}\right)$ 
	where 
	\begin{equation}
		V_{\pi,SRG} =
		\left\{
		\frac{\pi^2 \sigma_e (1-\sigma_e)}
		{c_1}
		+
		\frac{ (1-\pi)^2 \sigma_p (1-\sigma_p)}{c_2} 
		+ 
		\sum_{j=1}^k\frac{\gamma_j^2\rho_j(1-{\rho_j})}{c_3 s_{j}}
		\right\}
		(\sigma_e + \sigma_p -1)^{-2}.
		\label{eq:varpistd}
	\end{equation}
	The asymptotic variance $V_{\pi, SRG}$ can be consistently estimated by the plug-in estimator $\hat V_{\pi, SRG}$ defined by replacing $\sigma_e, \sigma_p, \rho_j, s_j, \pi$, and $c_l$ in \eqref{eq:varpistd} with $\hat \sigma_e, \hat \sigma_p, \hat \rho_j, n_{z_j} / n_3, \hat \pi_{SRG}$, and $n_l / n$ for $j = 1, \dots, k$ and $l = 1, 2, 3$.  Consistency of $\hat V_{\pi, SRG}$ holds by continuous mapping, and a proof of asymptotic normality and justification of \eqref{eq:varpistd} are in Appendix~\ref{sec:proofs_srg}. 
	
	standardization requires estimating the stratum-conditional mean of $X$, $\rho_j=P(X=1\mid Z=z_j)$. However, when $n_{z_j} = 0$ for some strata $j$, the corresponding estimator $\hat \rho_j$ is undefined, and $\hat \rho_{SRG}$ is then undefined as well. Values of $n_{z_j}$ may equal zero for two reasons. First, the study design may exclude these strata ($s_j=0$), a situation referred to as deterministic or structural nonpositivity \citep{westreich_invited_2010}. Second, even if $s_j > 0$, random nonpositivity can occur if no individuals with $Z=z_j$ are sampled, which may occur if $s_j$ is small or if $n_3$ is relatively small. When nonpositivity arises, an analytical approach often employed entails ``restriction'' \citep{westreich_invited_2010}, where the target population is redefined to consist only of strata $j$ for which $n_{z_j}>0$. However, this redefined target population may be less relevant from a public health or policy perspective.
	
	\subsection{Parametric standardization} 
	\label{sec:model}
	
Rather than redefining the target population, an alternative strategy for combatting positivity violations is to fit a parametric model to estimate all stratum-conditional means $\rho_j$. Such parametric models allow inference to the original target population and, when they are correctly specified, typically outperform nonparametric approaches \citep{petersen_diagnosing_2012, rudolph_parametric_2018, zivich_positivity_2022}. Assume the binary regression model $g(\rho_j)=\beta h(z_j)$ holds where $g$ is an appropriate link function for a binary outcome like the logit or probit function; $\beta$ is a row vector of $p$ regression coefficients with intercept $\beta_1$; and $h(z_j)$ is a user-specified $p$-vector function of the $j$th stratum's covariate values that may include main effects and interaction terms, with $l$th element denoted $h_l(z_j)$ and $h_1(z_j)$ set equal to one to correspond to an intercept. Let $\operatorname{supp}(z)$ be the covariate support in the sample, i.e., $\operatorname{supp}(z)=\{z_j: n_{z_j}>0\}$ with dimension $\dim\{\operatorname{supp}(z)\} =\sum_{j=1}^kI(n_{z_j}>0)$, and assume $p\leq \dim\{\operatorname{supp}(z)\}\leq k$. (Note that $\dim\{\operatorname{supp}(z)\}=k$ only when there is positivity, and in that case $\hat \pi_{SRG}$ can be used with no restriction needed.) 
	
	Under the assumed binary regression model, each $\rho_j$ is a function of the parameters $\beta$ and the covariates $z_j$ that define the $j$th stratum, denoted $\rho_j( \beta,z_j)=g^{-1}\{\beta h(z_j)\}$. A model-based standardized Rogan-Gladen estimator of $\pi$ is $\hat \pi_{SRGM}=(\hat \rho_{SRGM} + \hat \sigma_p - 1)/(\hat \sigma_e + \hat \sigma_p - 1)$, where $\hat \rho_{SRGM}=\sum_{j=1}^k \hat \rho_j (\hat \beta, z_j)\gamma_j$ and $\hat \beta$ is the MLE of $\beta$. Estimating equation theory can again be used to derive large-sample properties by replacing the $k$ equations for $\rho_1,\dots,\rho_k$ from Section~\ref{sec:standardization} with $p$ equations for $\beta_1,\dots,\beta_p$ corresponding to the score equations from the binary regression. 
	
	Let $\theta_m=(\sigma_e,\sigma_p,\beta_1,\dots,\beta_p,\rho,\pi)^T$ and $\hat \theta_m = (\hat \sigma_e, \hat \sigma_p, \hat \beta_1, \dots, \hat \beta_p, \hat \rho_{SRGM}, \hat \pi_{SRGM})^T$. The estimator $\hat \theta_m$ solves the vector $\sum \psi(X_i,Z_i; \delta_i,\theta_m)=\left(\sum \psi_e, \sum \psi_p, \sum \psi_\beta,  \psi_\rho,\psi_\pi\right)^T=0$ of estimating equations where $\sum \psi_e, \sum \psi_p$, and $\psi_\pi$ are as in Section~\ref{sec:methods_rg}; $\sum \psi_\beta$ is a $p$-vector with $j$th element $\sum \psi_{\beta_j}=\sum I(\delta_i=3)
	\left[
	X_i-g^{-1}\left\{\beta h(Z_i)\right\}
	\right]
	h_j(Z_i)$; and $\psi_\rho=\sum_{j=1}^k g^{-1}\{\beta h(Z_j)\}\gamma_j-\rho$. It follows that $\hat \theta_m$ is consistent and asymptotically normal and  $\sqrt{n}(\hat{\pi}_{SRGM} - \pi) \to_{d} \mathcal{N}\left(0, V_{\pi,SRGM}\right)$. The asymptotic variance $V_{\pi, SRGM}$ can be consistently estimated by $\hat V_{\pi, SRGM}$, the lower right element of the empirical sandwich variance estimator of the asymptotic variance of $\hat \theta_m$. A proof of asymptotic normality and the empirical sandwich variance estimator are given in Appendix~\ref{sec:proofs_srgm}. An R package for computing $\hat \pi_{SRG}$, $\hat\pi_{SRGM}$, and their corresponding variance estimators is available at \url{https://github.com/samrosin/rgStandardized}.
	
	\section{Simulation study}
	\label{sec:sims}
	
	Simulation studies were conducted to compare $\hat \pi_{RG}$, $\hat \pi_{SRG}$, and $\hat \pi_{SRGM}$. Four data generating processes (DGPs) were considered, within which different scenarios were defined through full factorial designs that varied simulation parameters $\pi, \sigma_e, \sigma_p, n_1, n_2$, and $n_3$. These DGPs featured no selection bias (DGP 1), selection bias with two strata (DGP 2), and more realistic selection bias with 40 strata and 80 strata (DGPs 3 and 4). 
	
	For each DGP and set of simulation parameters, sensitivity and specificity validation samples of size $n_1$ and $n_2$ were generated with $X$ distributed Bernoulli with a mean of $\sigma_e$ or $1-\sigma_p$, respectively. In DGPs 1 and 2 a main study of size $n_3$ was then generated where $Y$ was Bernoulli with mean $\pi$ and $X \mid Y$ was Bernoulli with mean $\sigma_eY + (1 - \sigma_p)(1 - Y)$; in DGPs 3 and 4 $X$ was generated from the distribution of $X \mid Z$, as described below. Simulation parameter values were selected based on the seroprevalence studies described in Section~\ref{sec:data_analysis}. Sensitivity was varied in $\sigma_e \in \{.8, .99\}$, specificity in $\sigma_p \in \{.8, .95, .99\}$, and prevalence in $\pi \in \{.01, .02, \dots, .20\}$. Sample sizes were $n_1 = 40$, $n_2 = 250$, and $n_3 = 2500$. The full factorial design led to 120 scenarios per DGP, and within each scenario 1,000 simulations were conducted unless otherwise specified. Performance was measured by: (a) mean bias, computed as the mean of $\hat \pi - \pi$ for each estimator $\hat \pi$; (b) empirical coverage, i.e., whether the 95\% Wald-type CIs based on each variance estimator $\hat V_{\pi}$ contained the true prevalence; (c) mean squared error (MSE), computed as the mean of  $(\hat \pi - \pi)^2$ for each estimator $\hat \pi$. R code implementing the simulations is available at \url{https://github.com/samrosin/rgStandardized_ms}.
	
	\subsection{No selection bias}
	\label{sec:sims_rg}
	
		For DGP 1, 10,000 simulations were conducted to assess the performance of $\hat \pi_{RG}$ when no selection bias was present. The estimator $\hat \pi_{RG}$ was generally unbiased, as seen in Appendix Figure~\ref{fig:dgp1_bias}. Performance improved as $\sigma_e$ and $\sigma_p$ tended toward 1, with $\sigma_p$ being a stronger determinant of bias. An exception to these results occurred when $\pi \leq 0.05$ and $\sigma_p \leq 0.95$, in which case $\hat \pi_{RG}$ overestimated the true prevalence. The Rogan-Gladen estimator without truncation was also evaluated in this DGP to determine if truncation caused the bias. While the non-truncated estimator was slightly biased, the magnitude of the bias was less than 0.002 in all scenarios, suggesting the bias of $\hat \pi_{RG}$ in low prevalence, low specificity settings is due largely to truncation.
	
	Wald CIs based on $\hat V_{\pi, RG}$ attained nominal coverage in almost every scenario, as seen in Appendix Figure~\ref{fig:dgp1_coverage}. However, when some parameters were near their boundaries, coverage did not reach the nominal level. For instance, when $\pi$ was 0.01 and $\sigma_p$ was 0.99, 95\% CIs covered in 90\% and 91\% of simulations for two values of $\sigma_e$. These variable CI coverage results concord with previous simulation studies evaluating $\hat V_{\pi, RG}$ \citep{lang_confidence_2014}. The MSE of $\hat \pi_{RG}$, shown in Appendix Figure~\ref{fig:dgp1_mse}, tended to increase with $\pi$ and decrease as $\sigma_e$ and $\sigma_p$ approached 1.
	
	\subsection{Low-dimensional selection bias}
	\label{sec:dgp2}
	
	In DGP 2, the target population was comprised of two strata defined by a covariate $Z \in \{z_1, z_2\}$ with proportions $\gamma_1=\gamma_2=.5$. Within the main study, $Z$ was generated from a binomial distribution of sample size $n_3$ and sampling probabilities $(.2, .8)$. Individuals' serostatuses were generated from the conditional distribution $Y \mid Z$, which was such that $P(Y=1 \mid Z=z_1)=1.5 \pi$ and $P(Y=1 \mid Z=z_2)= 0.5 \pi$ for each value of $\pi$. In each simulation $\hat \pi_{RG}$ and $\hat \pi_{SRG}$ and their corresponding 95\% CIs were computed.
	
	The nonparametric standardized estimator $\hat \pi_{SRG}$ was empirically unbiased for true prevalences $\pi \geq 0.05$, as seen in Appendix Figure~\ref{fig:dgp2_bias}, and 95\% CIs based on $\hat V_{\pi, SRG}$ attained nominal coverage in almost every scenario, as seen in Appendix Figure~\ref{fig:dgp2_coverage}. As with $\hat \pi_{RG}$ in DGP 1, CI coverage for $\hat \pi_{SRG}$ was slightly less than the nominal level for very low $\pi$ and for $\sigma_p$ near the boundary, e.g., coverage was 91\% for $\pi=.01$ and $\sigma_e=\sigma_p=.99$. MSE trends for $\hat \pi_{SRG}$ were similar to those of $\hat \pi_{RG}$ in DGP 1, as seen in Appendix Figure~\ref{fig:dgp2_mse}. Appendix Figures~\ref{fig:dgp2_bias}, \ref{fig:dgp2_coverage}, and \ref{fig:dgp2_mse} show that $\hat \pi_{RG}$ performed poorly under selection bias, with large negative bias, CI coverage far less than the nominal level in most cases, and much greater MSE than $\hat \pi_{SRG}$.

	\subsection{More realistic selection bias } \label{sec:sims_realistic}
	
		\subsubsection{DGP 3} \label{sec:dgp3}
	
\begin{figure}[!ht]
	\centering
	\includegraphics[width=1\textwidth]{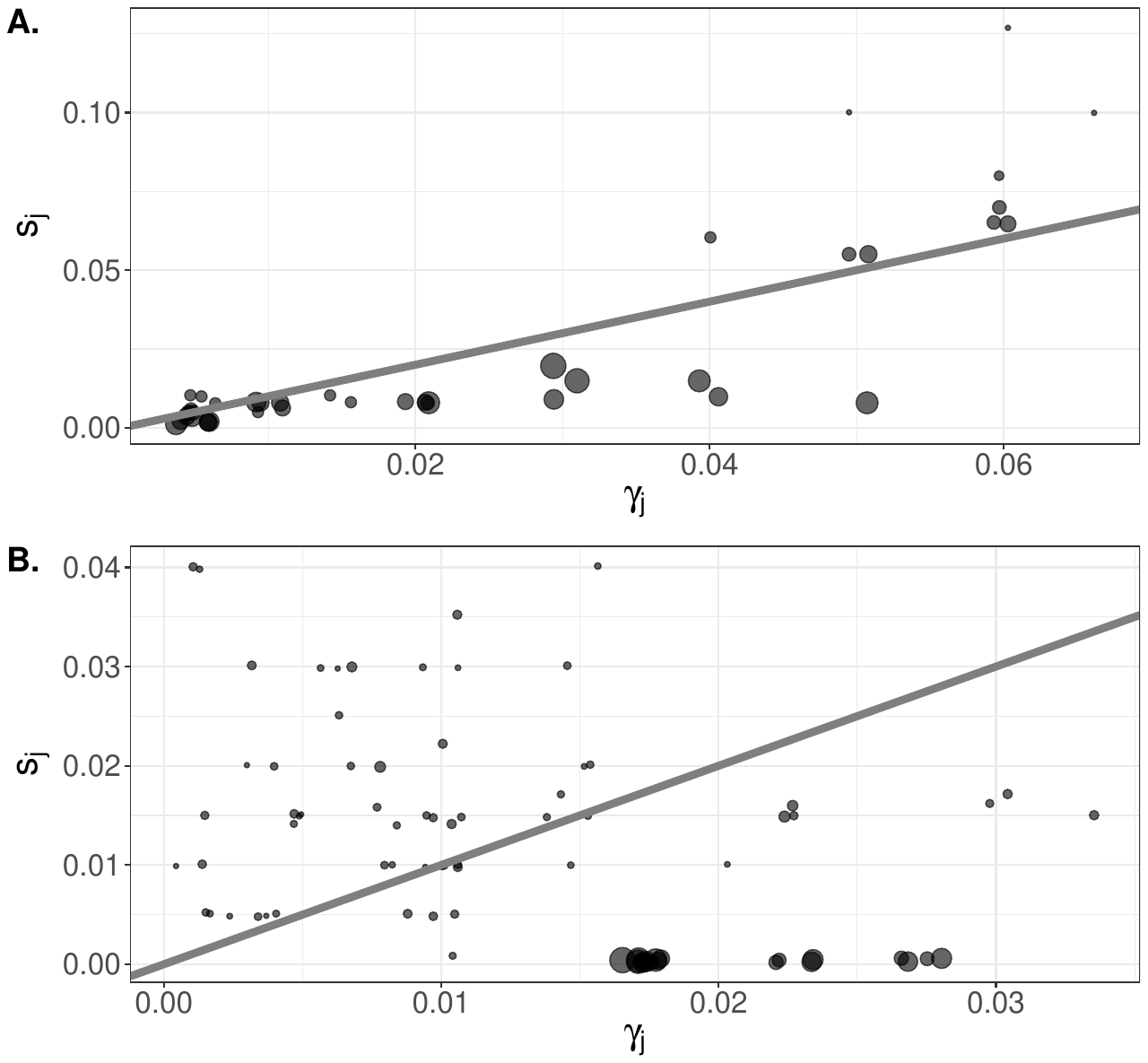}   
	\caption{Panels A and B represent selection bias in the simulation studies of DGPs 3 and 4, described in Sections~\ref{sec:dgp3} and \ref{sec:dgp4}, respectively. Circle size is proportional to prevalence. Points are jittered slightly for legibility, and the diagonal lines denote equality between $\gamma_j$ (stratum proportion) and $s_j$ (sampling probability).}
	\label{fig:dgp3_selectionbias}
\end{figure}

DGPs 3 and 4 compared $\hat \pi_{SRG}$ and $\hat \pi_{SRGM}$ in scenarios with larger numbers of strata. In DGP 3, three covariates were defined as $Z_{1}\in\{z_{10},z_{11}\}$, $Z_{2}\in \{z_{20},z_{21},z_{22},z_{23}\}$, and $Z_{3}\in\{z_{30},z_{31},z_{32},z_{33},z_{34}\}$, leading to $k=40$ strata with proportions $(\gamma_1, \dots, \gamma_{40})$. Within the main study, $Z$ was generated as multinomial with size $n_3$ and known sampling probabilities. Figure~\ref{fig:dgp3_selectionbias}(a) shows the structure of selection bias in DGP 3 by comparing the stratum proportions and sampling probabilities. Some low-prevalence strata that frequently occur in the population were oversampled, while most remaining strata were undersampled. Individuals' test results were generated from the conditional distribution $X \mid Z$, where
\begin{equation*}
	\begin{split}
		\operatorname{logit}\{P(X=1\mid Z)\} &= \beta_0+\beta_{1}I(Z_1=z_{11})+\beta_{2}I(Z_2=z_{20})+\beta_{3}I(Z_2=z_{21}) \\ 
		&+ \beta_4I(Z_3=z_{30})+\beta_5I(Z_3=z_{31}).
	\end{split}
\end{equation*}
The parameters $\beta_1=-1, \ \beta_2 = -.6, \ \beta_3 = .8, \ \beta_4 = .6,$ and  $\beta_5 = .4$ were set to reflect differential prevalences by stratum, while a ``balancing intercept'' $\beta_0$ \citep{rudolph_simulation_2021} was set to different values so that $\pi$ equalled (approximately) $\{.01, .02, \dots, .20\}$. The nonparametric estimator $\hat \pi_{SRG}$ and corresponding CI were computed using a restricted target population when random nonpositivity arose; the values of $\pi$ used to compute bias and coverage were based on the total (unrestricted) population, which is the parameter of interest. The parametric estimator $\hat \pi_{SRGM}$ was computed with a correctly-specified logistic regression model, with parameters estimated using maximum likelihood. 

\begin{figure}[!ht]
	\centering
	\includegraphics[width=1\textwidth]{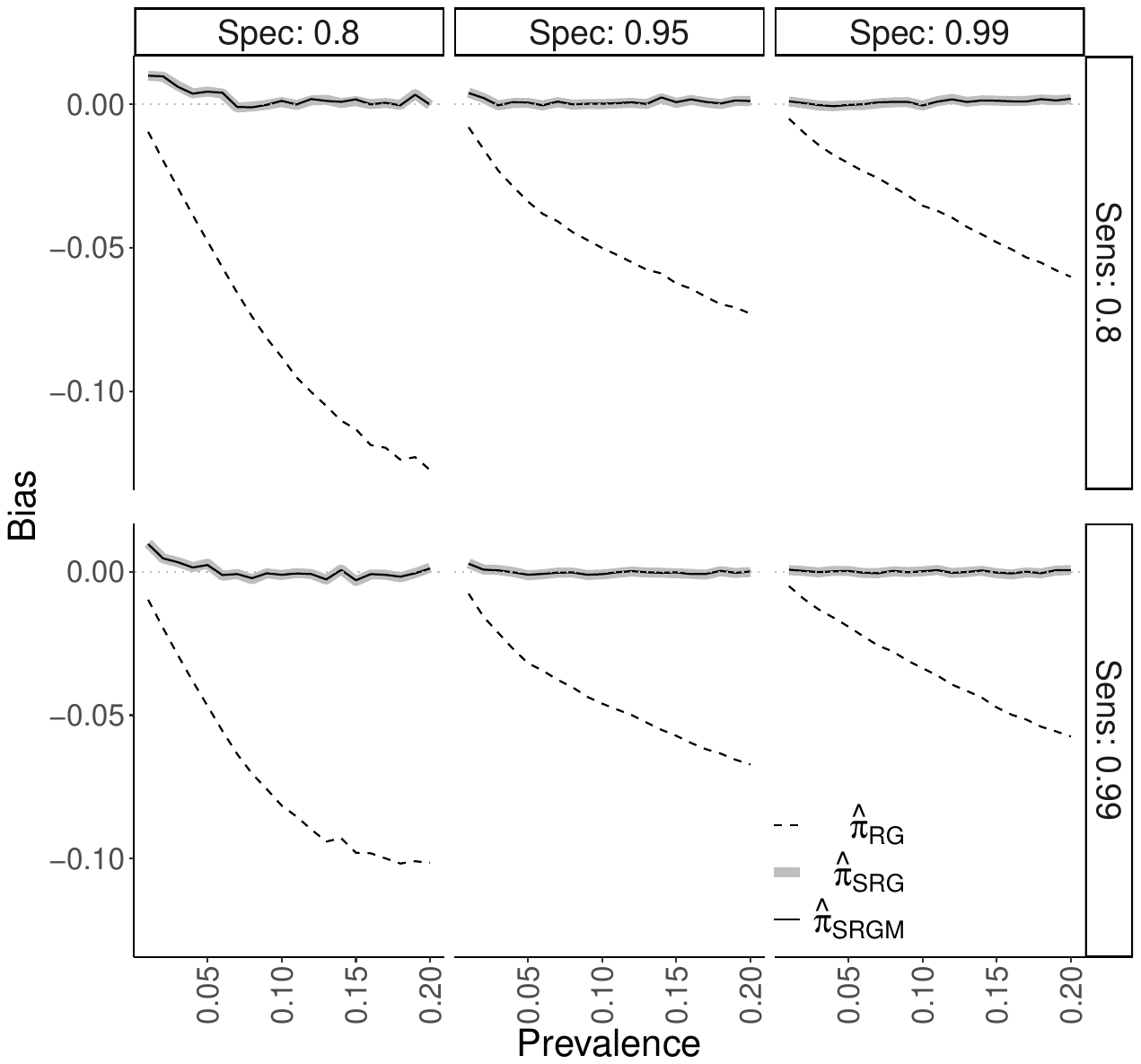}
	\caption{Empirical bias of the Rogan-Gladen ($\hat \pi_{RG}$), nonparametric standardized ($\hat \pi_{SRG}$), and logistic regression standardized ($\hat \pi_{SRGM}$) estimators from simulation study for DGP 3, described in Section~\ref{sec:dgp3}. The six facets correspond to a given combination of sensitivity (`Sens') and specificity (`Spec'). } 
	\label{fig:dgp3_bias}
\end{figure}

Both $\hat \pi_{SRG}$ and $\hat \pi_{SRGM}$ performed well in this scenario. Figure~\ref{fig:dgp3_bias} shows that the estimators were generally empirically unbiased, though modest bias occurred when $\sigma_p=0.8$ and $\pi$ was low. As in DGP 2, $\hat \pi_{RG}$ exhibited substantial bias and the CIs based on $\hat \pi_{RG}$ did not attain nominal coverage. Appendix Figure~\ref{fig:dgp3_coverage} shows 95\% CIs based on either $\hat V_{\pi, SRG}$ or $\hat V_{\pi, SRGM}$ attained nominal coverage, with slight under-coverage for $\pi < 0.05$, similar to the results from DGPs 1 and 2. For $\pi = .01$ and $\sigma_p=.99$, coverage was 92\% and 90\% based on $\hat V_{\pi,SRG}$ and 91\% and 90\% based on $\hat V_{\pi,SRGM}$ for $\sigma_e \in \{.8, .99\}$, respectively. The two standardized estimators had roughly equivalent MSE (Appendix Figure~\ref{fig:dgp3_mse}). On average across all 120 scenarios, positivity was present in 89\% (range of 86\%-92\%) of simulated datasets, i.e., these datasets included all strata in the target population. 

	\subsubsection{DGP 4} \label{sec:dgp4} 
	
	Data were generated as in DGP 3, but the inclusion of a fourth covariate $Z_4 \in \{z_{40}, z_{41}\}$ led to 80 strata. The conditional distribution $X \mid Z$ was such that $\operatorname{logit}\{P(X = 1 \mid Z)\} = \nu h(Z)$, where $h(Z)$ here contains the same terms as in DGP 3 plus a main effect for $I(Z_4 = z_{41})$ with corresponding coefficient $\nu_6$.  Regression parameters were a balancing intercept $\nu_0$, $\nu_1 = -1$, $\nu_2 = 3.25$, $\nu_3 = .8$, $\nu_4 = .6$, $\nu_5 = .4$, and $\nu_6 = .1$. The larger value for $\nu_2$, as compared to $\beta_2$, led to a stronger relationship between $X$ and $Z$ than was present in DGP 3. Figure~\ref{fig:dgp3_selectionbias}(b) displays selection bias in DGP 4. Some of the highest-prevalence and most commonly-occurring strata were undersampled to a greater degree than occurred in DGP 3, so in this sense there was more selection bias in DGP 4. The parametric estimator $\hat \pi_{SRGM}$ was again computed with a correctly-specified logistic regression model using maximum likelihood for parameter estimation.
	
\begin{figure}[!ht]
		\centering
		\includegraphics[width=1\textwidth]{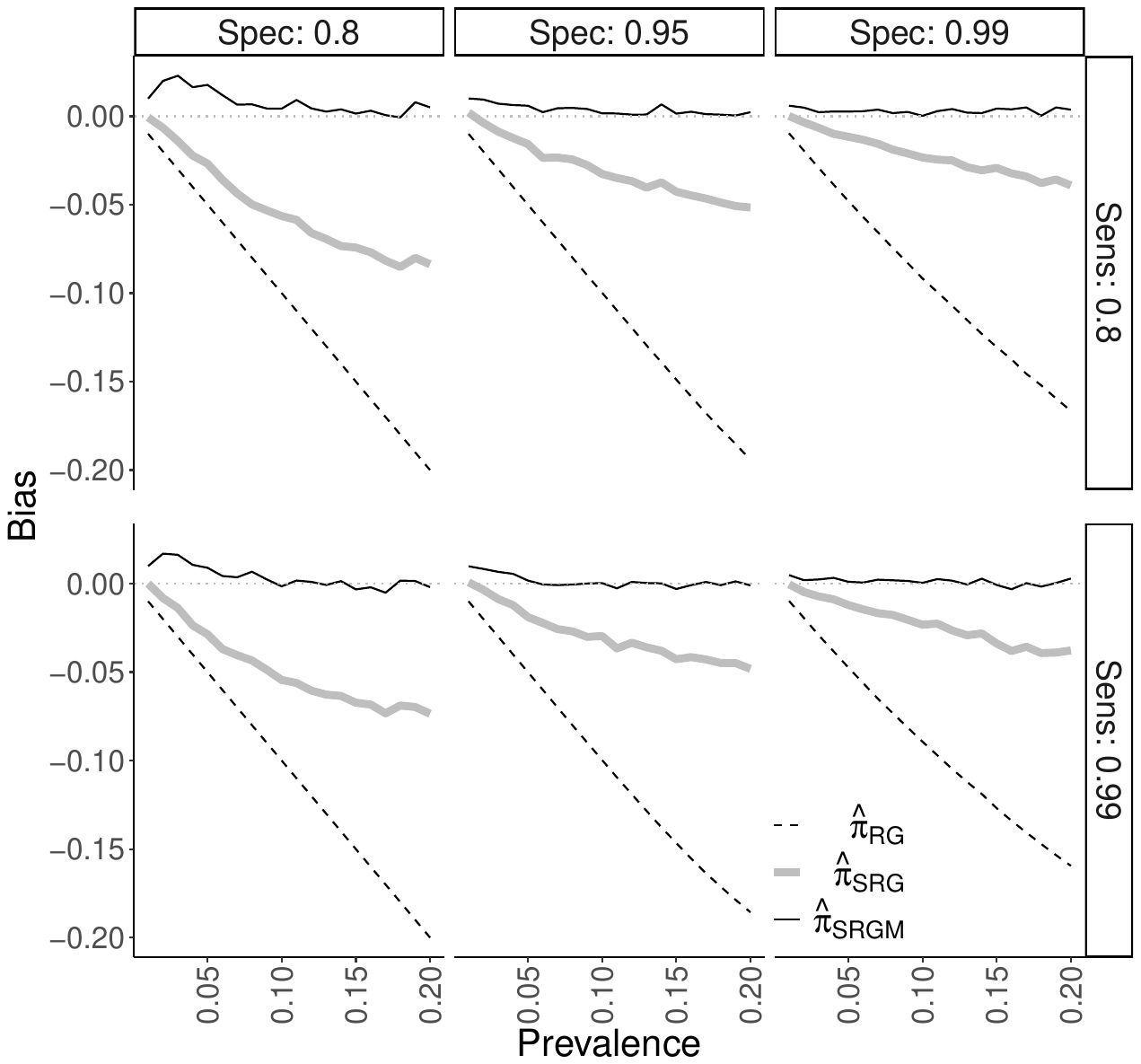}
		\caption{Bias results from simulation study on DGP 4, described in Section~\ref{sec:dgp4}. Figure layout is as in Figure~\ref{fig:dgp3_bias}.}
		\label{fig:dgp4_bias}
	\end{figure}

	Results for the DGP 4 simulations are shown in Figure~\ref{fig:dgp4_bias} and Appendix Figures~\ref{fig:dgp4_coverage}-\ref{fig:dgp4_mse}. Figure~\ref{fig:dgp4_bias} shows that only $\hat \pi_{SRGM}$ was generally unbiased under DGP 4, although there was positive bias when $\sigma_p = .8$ and $\pi < .1$. The nonparametric $\hat \pi_{SRG}$ typically had a moderately negative bias. Nonpositivity almost always occurred (in either all or all but one of the simulations, for each of the 120 scenarios). The worse bias of $\hat \pi_{SRG}$ may be explained by restriction leading to bias under nonpositivity. CIs based on $\hat V_{\pi, SRGM}$ typically attained nominal or close-to-nominal coverage, unlike those based on $\hat V_{\pi, SRG}$ or $\hat V_{\pi, RG}$, as seen in Appendix Figure~\ref{fig:dgp4_coverage}. For instance, when $\sigma_p = 0.8$, the lowest coverage for CIs based on $\hat V_{\pi, SRGM}$ was 92\% across all 40 combinations of $\sigma_e$ and $\pi$. However, the $\hat V_{\pi, SRGM}$ based CIs exhibited undercoverage when $\sigma_p = .99$ and prevalence $\pi$ was low. For example, coverage of these CIs was only 59\% when $\pi = 0.01$ and $\sigma_e=\sigma_p=.99$; Appendix Figure~\ref{fig:dgp4_diagnostics} shows this undercoverage is due to 39\% of the $\hat \pi_{SRGM}$ estimates being truncated to 0 with corresponding CIs which were overly narrow. Note that $\hat V_{\pi, SRGM}$ was negative for two simulations in a single scenario, and these ``Heywood cases'' \citep{kolenikov_testing_2012} were ignored in the coverage calculation for that scenario. The MSE of $\hat \pi_{SRGM}$ tended to be less than that of $\hat \pi_{SRG}$ (Appendix Figure~\ref{fig:dgp4_mse}).
	
	In summary, both the nonparametric and parametric standardized estimators $\hat \pi_{SRG}$ and $\hat \pi_{SRGM}$ had low empirical bias and close to nominal 95\% CI coverage when there was positivity or near positivity. As the number of covariates, amount of selection bias, and potential for nonpositivity increased, the (correctly-specified) parametric $\hat \pi_{SRGM}$ generally maintained its performance while $\hat \pi_{SRG}$ had greater empirical bias and the intervals based on $\hat V_{\pi, SRG}$ did not attain nominal coverage levels.
	
	\subsection{Model misspecification}
	\label{sec:sims_misspec}
	
	The performance of $\hat \pi_{SRGM}$ was assessed in scenarios similar to DGPs 3 and 4, but under model misspecification. Here, the true conditional distributions of $Y \mid Z$ were $\operatorname{logit}\{P(Y = 1 \mid Z)\} = \beta h(Z)$ and $\operatorname{logit}\{P(Y = 1 \mid Z)\} = \nu h(Z)$, where $\beta h(Z)$ and $\nu h(Z)$ are the specifications used in the models for $\operatorname{logit}\{P(X = 1 \mid Z)\}$ in DGPs 3 and 4, respectively. The test results $X$ were generated from $X \mid Y$ as in DGPs 1 and 2. The results shown in Appendix Figures~\ref{fig:dgp5}-\ref{fig:dgp6_mse} demonstrate that, in terms of bias, 95\% CI coverage, and MSE, inferences based on $\hat \pi_{SRGM}$ were generally robust to this misspecification. For DGP 3 the true $X \mid Z$ distribution was $\Pr(X = 1 \mid Z) = [\operatorname{logit}^{-1}\{\beta h(Z)\} + \sigma_p - 1] / [\sigma_e + \sigma_p - 1]$, while the model-based estimator incorrectly assumed $\Pr(X = 1 \mid Z)=\operatorname{logit}^{-1}\{\beta h(Z)\}$, and likewise for DGP 4 with $\nu h(Z)$ replacing $\beta h(Z)$. Thus the degree of misspecification was determined by $\sigma_e$ and $\sigma_p$, with values farther from 1 leading to greater misspecification. For all simulation scenarios $\sigma_e \geq 0.8$ and $\sigma_p \geq 0.8$ such that the overall degree of misspecification was generally mild, potentially explaining the robustness of $\hat \pi_{SRGM}$ to model misspecification in these simulations.
	
Robustness of the model-based estimator $\hat \pi_{SRGM}$ was also assessed when the model was misspecified by omitting a variable. Under DGP 3, $\hat \pi_{SRGM}$ was estimated based on three misspecified logistic regression models, each omitting one of the three variables $Z_1$, $Z_2$, or $Z_3$ (i.e., omitting all indicator variables that included the variable). Results displayed in Appendix Figure~\ref{fig:dgp10} show the empirical bias of $\hat \pi_{SRGM}$ depended on which variable was omitted, with substantial bias when $Z_2$ was not included in the model. These results demonstrate that $\hat \pi_{SRGM}$ may not be robust to model misspecification due to variable omission.

	\section{Applications} 
	\label{sec:data_analysis}
	
	\subsection{NYC seroprevalence study}
	\label{sec:nyc}
	The methods were applied to a seroprevalence study in New York City (NYC) that sampled patients at Mount Sinai Hospital from February 9 to July 5, 2020 \citep{stadlbauer_repeated_2021}. Patients were sampled from two groups: (1) a `routine care' group visiting the hospital for reasons unrelated to COVID-19, including obstetric, gynaecologic, oncologic, surgical, outpatient, cardiologic, and other regular visits; (2) an `urgent care' group of patients seen in the emergency department or admitted to the hospital for urgent care. Analyses were stratified by these two care groups. The urgent care group may have included individuals seeking care for moderate-to-severe COVID-19 \citep{stadlbauer_repeated_2021}; this would potentially violate the assumption of equal sampling probabilities within strata, but standardized analysis of the urgent care group is included here to demonstrate the methods. The routine care group was thought to be more similar to the general population \citep{stadlbauer_repeated_2021}. Serostatus was assessed using a two-step enzyme-linked immunosorbent assay (ELISA) with estimated sensitivity of $\hat \sigma_e = 0.95$ from $n_1 = 40$ PCR-confirmed positive samples and estimated specificity of $\hat \sigma_p = 1$ from $n_2 = 74$ negative controls, 56 of which were pre-pandemic and 18 of which did not have confirmed SARS-CoV-2 infection. 
	

In this analysis, the samples were grouped into five collection rounds of approximately equal length of 	time. The demographics considered were sex, age group, and race. Sex was categorized as male/female, with one individual of indeterminate sex excluded. Five age groups were $[0, 20), [20, 40), [40, 60), [60, 80)$, and $[80, 103]$. Race was coded as Asian, Black or African American, Other, and White, with 446 individuals of unknown race excluded. After exclusions, the sample size ranged from $n_3 = 937$ to 1576 in the routine care group and $n_3 = 622$ to 955 in the urgent care group across the collection rounds. The target population for standardization was NYC (8,336,044 persons), with stratum proportions and population size obtained from the 2019 American Community Survey \citep{us_census_bureau_american_2019}. Table~\ref{tab:nyc} compares the distributions of sex, age group, and race in the routine and urgent care groups to the NYC population. Women were overrepresented in the routine care group relative to the general population of NYC. Persons aged 0-19 were underrepresented in both groups, and persons aged 60 and older were overrepresented. Persons with race classified as Other were overrepresented in both groups relative to the NYC population. There was nonpositivity in 4/5 and 5/5 collection rounds for the routine and urgent care groups, respectively, and $\hat \pi_{SRG}$ made inference to restricted target populations. The model-based estimators $\hat \pi_{SRGM}$ included main effects for sex, age group, and race and an interaction term between sex and age group.  

\begin{table}
\caption{\label{tab:nyc}Demographic comparisons of the New York City (NYC) seroprevalence study (Feb 9 - July 5, 2020) routine and urgent care group samples with the NYC population. Data on the NYC population are from the 2019 American Community Survey\citep{us_census_bureau_american_2019}. Sample size is denoted by $n$. Some column totals do not sum to 100\% because of rounding.}
\begin{tabular}{llcccccc} \hline
	&
	&
	\multicolumn{2}{c}{Routine care}& \multicolumn{2}{c}{Urgent care}& \multicolumn{2}{c}{NYC}  \\ 
	
	& & $n$ & \% & $n$ & \% & $n$ & \% \\ \hline
	
	& & 6,348 & 100 & 3,898 & 100 & 8,336,044 & 100 \\ \hline
	
	\multirow{2}{*}{Sex} & Female & 4,274 & 67 & 1,789& 46 & 4,349,715& 52\\
	
	& Male & 2,074 & 33 &2,109 &54 & 3,986,329& 48 \\ \hline
	
	\multirow{5}{*}{Age} & 0-19 & 238 & 4 & 93 & 2 & 1,887,268 & 23 \\ 
	
	& 20-39 &2,624 & 41& 551& 14& 2,608,394
	& 31 \\
	
	& 40-59 & 1,396&22 &1,065 &27 & 2,080,599 & 25 \\ 
	
	& 60-79 &1,780& 28& 1,633& 42  & 1,426,301& 17 \\ 
	
	& 80+ & 310& 5& 556& 14& 333,482
	& 4 \\ \hline
	
	\multirow{4}{*}{Race} & Asian & 562 & 9& 217& 6& 1,202,530 & 14 \\
	
	& 	Black or African-American & 1,287&20 & 1,051& 27 & 2,057,795& 25 \\
	
	& 	Other& 1,774&28 & 1,514 & 39& 1,528,503& 18 \\
	
	& White  &2,725 & 43& 1,116 & 29 & 3,547,216
	& 43
	
\end{tabular}
\end{table}

\begin{figure}[!ht]
\centering
\includegraphics[width=1\textwidth]{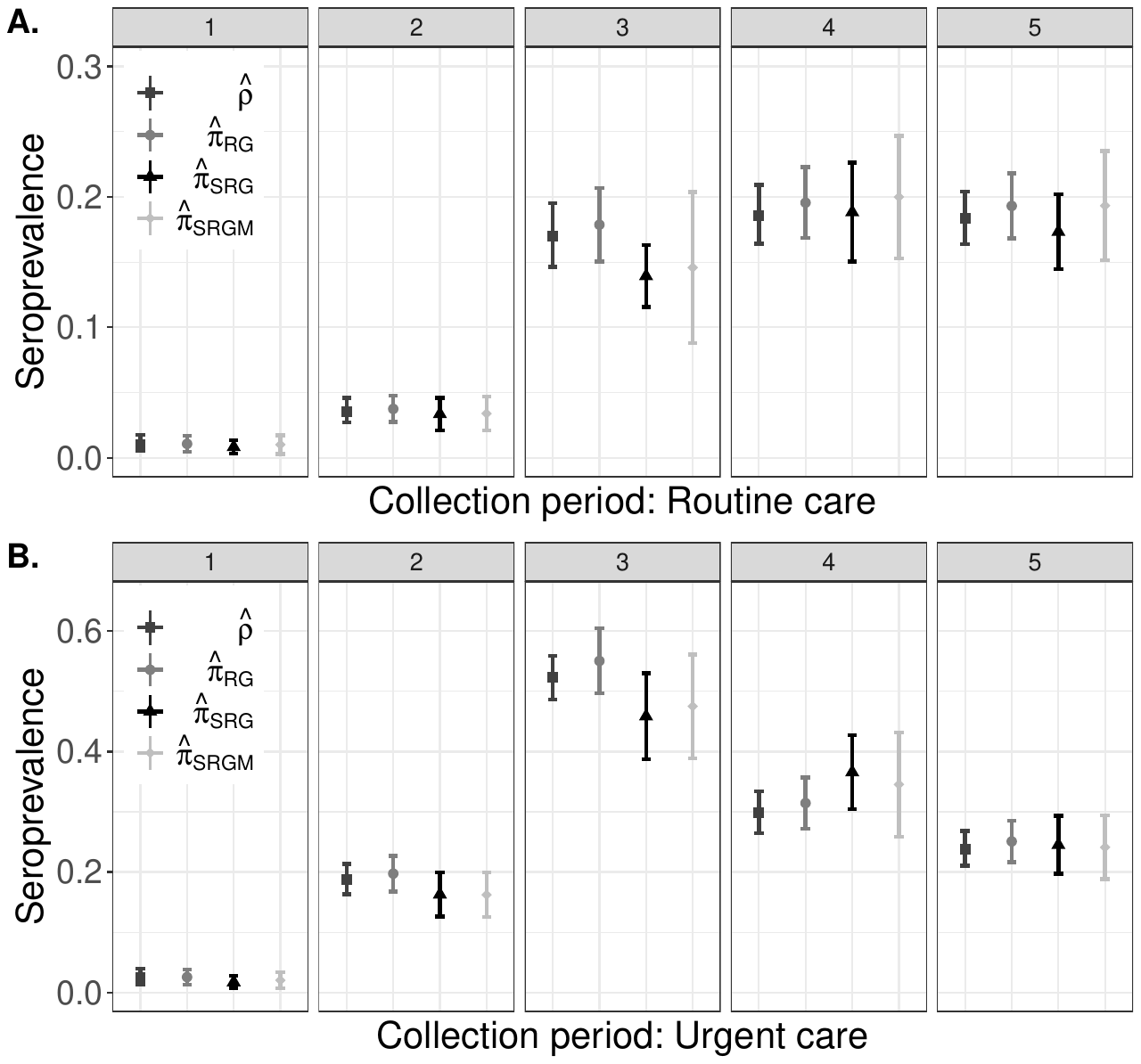}
\caption{Estimates and corresponding 95\% confidence intervals for each of five collection rounds for the NYC seroprevalence study \citep{stadlbauer_repeated_2021}, stratified by routine and urgent care groups, described in Section~\ref{sec:nyc}.}
\label{fig:nyc}
\end{figure}

Seroprevalence estimates are presented in Figure~\ref{fig:nyc}. Adjusting for assay sensitivity and specificity resulted in slightly higher estimates and slightly wider CIs than the naive estimator $\hat \rho$. standardization had the largest impact on the estimates in the third round, when $\hat \pi_{RG}$ and the standardized estimators differed by as much as 9 percentage points. The standardized estimates were accompanied by wider CIs relative to $\hat \rho$ and $\hat \pi_{RG}$, reflecting greater uncertainty associated with estimating seroprevalence when not assuming the main study data constitute a random sample from the target population.
	
\subsection{Belgium seroprevalence study}
\label{sec:belgium}

The standardized Rogan-Gladen methods were applied to a nationwide SARS-CoV-2 seroprevalence study in Belgium conducted across seven week-long collection rounds between March and October 2020 \citep{herzog_seroprevalence_2022}. The final collection round took place before the first vaccine authorization in the European Union in December 2020. Residual sera were collected in a stratified random sample from private laboratories encompassing a wide geographical network, with stratification by age group (10 year groups from 0-9, 10-19, \dots, 90-plus), sex (male or female), and region (Wallonia, Flanders, or Brussels). The presence of SARS-CoV-2 IgG antibodies was determined using a semi-quantitative EuroImmun ELISA. Based on validation studies of $n_1 = 181$ reverse transcription PCR-confirmed COVID-19 cases and $n_2 = 326$ pre-pandemic negative controls, sensitivity and specificity were estimated to be $\hat \sigma_e = .851$ and $\hat \sigma_p = .988$ \citep[][Table S1.1]{herzog_seroprevalence_2022}. The number of samples for assessing seroprevalence varied between $n_3 = $ 2,960 and $n_3 = $ 3,910 across the seven collection rounds. 

In this analysis, nationwide seroprevalence in Belgium was estimated during each collection round standardized by age group, sex, and province (11 total), using 2020 stratum proportion data from the Belgian \citet{federal_planning_bureau_federal_2021}. Province was used rather than region to match the covariates selected for weighting in \citet{herzog_seroprevalence_2022}. Table~\ref{tab:belgium} compares the sex, age group, and province distributions in collection rounds 1 and 7 to the Belgian population as a whole. The seroprevalence study samples were similar to the population, although the study overrepresented older persons and underrepresented younger persons relative to the population. In six of the seven collection rounds nonpositivity arose, with between 2 to 15 of the 220 strata not sampled, so restricted target populations were used for computation of $\hat \pi_{SRG}$. For $\hat \pi_{SRGM}$, each logistic regression model had main effects for age group, sex, and province, as well as an interaction term between age group and sex. 

\begin{table}
	\caption{\label{tab:belgium}Demographic comparisons of the 2020 Belgium seroprevalence study sample participants in collection rounds 1 (30 March - 5 Apr) and 7 (9 - 12 Sept) with the Belgium population. Data on the Belgium population are from the \citet{federal_planning_bureau_federal_2021}. Sample size is denoted by $n$. Some column totals do not sum to 100\% because of rounding.}
	\begin{tabular}{llcccccc} \hline
		&
		&
		\multicolumn{2}{c}{Round 1}& \multicolumn{2}{c}{Round 7}& \multicolumn{2}{c}{Belgium}  \\ 
		
		& & $n$ & \% & $n$ & \% & $n$ & \% \\ \hline
		
		& & 3,910 & 100 & 2,966 & 100 & 11,492,641 & 100 \\ \hline
		
		\multirow{2}{*}{Sex} & Female & 2,111 & 54 & 1,589 & 54 & 5,832,577 & 51\\
		
		& Male & 1,799 & 46 &1,377 & 46 & 5,660,064& 49 \\ \hline
		
		\multirow{10}{*}{Age} & 0-9 & 36 & 1 & 68 & 2 & 1,269,068 & 11 \\ 
		
		& 10-19 & 294 & 8 & 405 & 14& 1,300,254 & 11 \\
		
		& 20-29 & 436 & 11 & 402 & 14  & 1,407,645 & 12 \\
		
		& 30-39 & 461 & 12 & 397 & 13 & 1,492,290 & 13 \\ 
		
		& 40-49 & 468 & 12 & 397 & 13 & 1,504,539 & 13 \\ 
		
		& 50-59 & 498 & 13 & 400 & 13 & 1,590,628 & 14 \\ 
		
		& 60-69 & 507 & 13 & 406 & 14 & 1,347,139 & 12 \\ 
		
		& 70-79 & 506 & 13 & 204 & 7 & 924,291 & 8 \\ 
		
		& 80-89 & 493 & 13 & 160 & 5 & 539,390 & 5 \\ 
		
		& 90+ & 211 & 5 & 127 & 4 & 117,397 & 1 \\ \hline
		
		\multirow{11}{*}{Province} & Antwerp & 819 & 21 & 473 & 16 & 1,869,730 & 16 \\
		
		& Brussels & 204 & 5 & 288 & 10 & 1,218,255 & 11 \\
		
		& East Flanders & 388 & 10 & 392 & 13 & 1,525,255 & 13 \\
		
		& Flemish Brabant  & 261 & 7 & 317 & 11 & 1,155,843 & 10 \\
		
		& Hainaut & 245 & 6 & 271 & 9 & 1,346,840 & 12 \\
		
		& Liege & 515 & 13 & 425 & 14 & 1,109,800 & 10 \\
		
		& Limburg & 318 & 8 & 280 & 9 & 877,370 & 8\\
		
		& Luxembourg & 254 & 7 & 177 & 6 & 286,752 & 3 \\
		
		& Namur & 352 & 9 & 170 & 6 & 495,832 & 4 \\
		
		& Walloon Brabant & 145 & 4 & 101 & 3 & 406,019 & 4 \\
		
		& West Flanders & 409 & 10 & 72 & 2 & 1,200,945 & 10 
	\end{tabular}
\end{table}

\begin{figure}[!ht]
	\centering
	\includegraphics[width=1\textwidth]{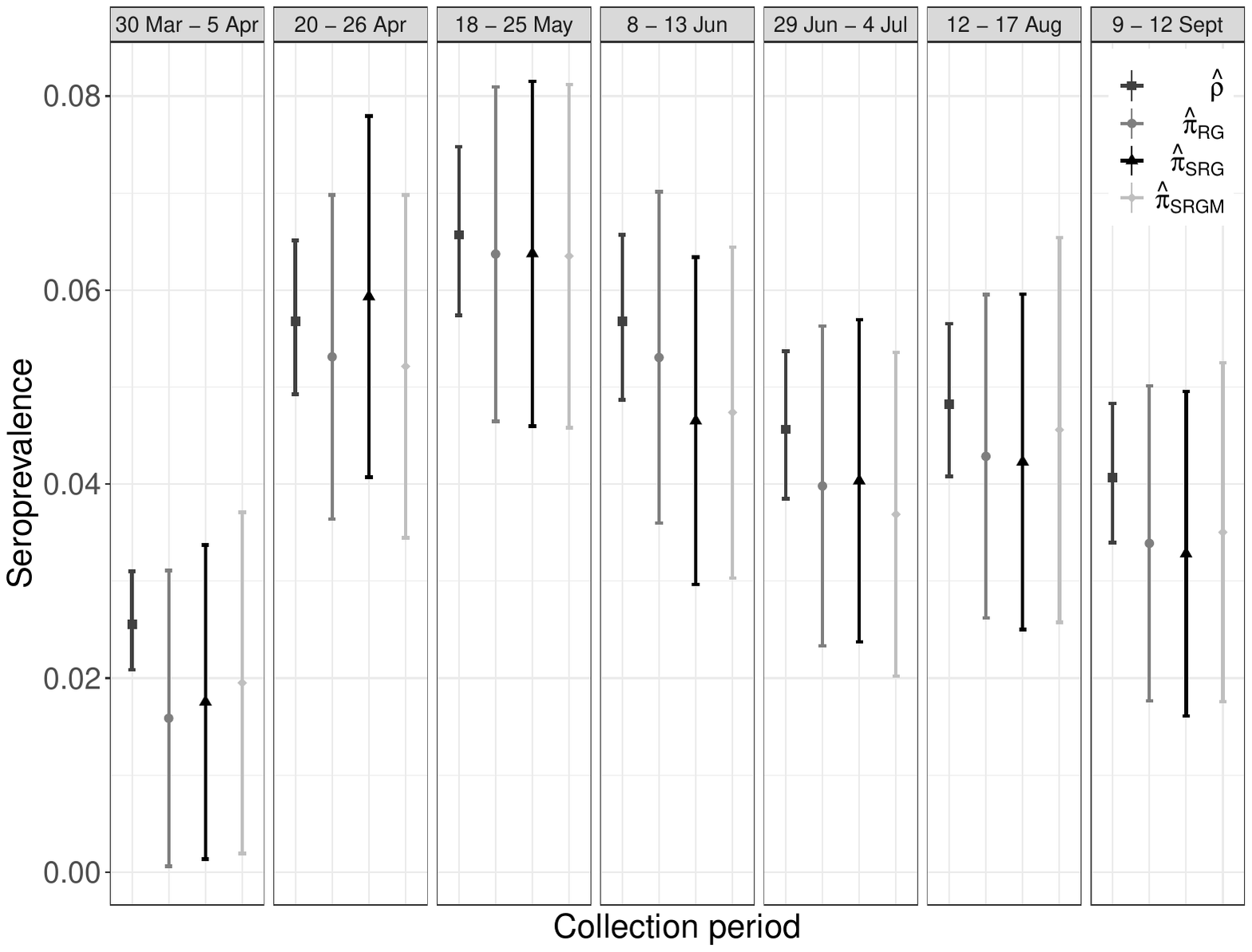}
	\caption{	Estimates and corresponding 95\% confidence intervals for each of seven collection rounds for the 2020 Belgian seroprevalence study \citep{herzog_seroprevalence_2022}, described in Section~\ref{sec:belgium}.}
	\label{fig:forest}
\end{figure}

Figure~\ref{fig:forest} displays point estimates and CIs for $\hat \pi_{RG}$, $\hat \pi_{SRG}$, and $\hat \pi_{SRGM}$ alongside those for the unadjusted, or naive, sample prevalence $\hat \rho$ for each collection round (with exact Clopper-Pearson 95\% CIs). The naive estimates $\hat \rho$ were typically greater than the other three estimates and had narrower CIs. The greatest differences were between $\hat \rho$ and $\hat \pi_{RG}$, which can be attributed to (estimated) measurement error in the assay. Both standardized estimates $\hat \pi_{SRG}$ and $\hat \pi_{SRGM}$ were similar in value to $\hat \pi_{RG}$ in most collection periods. These estimates, in combination with the stratified random sampling design, suggest that the magnitude of measurement error in this study may have been larger than that of selection bias.

\subsection{North Carolina seroprevalence study}
\label{sec:screennc}
The standardization methods of Section~\ref{sec:selection_bias} were also applied to ScreenNC, which tested a convenience sample of $n_3 = $ 2,973 asymptomatic patients age 20 and older in North Carolina (NC) for antibodies to SARS-CoV-2 between April to June 2020 \citep{barzin_sars-cov-2_2020}, before the authorization of vaccines in the United States. These patients were seeking unrelated medical care at eleven sites in NC associated with the University of North Carolina (UNC) Health Network. The presence of antibodies was determined with the Abbott Architect SARS-CoV-2 IgG assay. Based on validation studies of $n_1 = 40$ reverse transcription PCR confirmed positive patients and $n_2 = 277$ pre-pandemic serum samples assumed to be negative, sensitivity was estimated as $\hat \sigma_e = 1$ and specificity as $\hat \sigma_p = 0.989$. 

In our analysis, seroprevalence was estimated in two relevant target populations. First, standardization was made to the population patients accessing the UNC Health Network during a similar timeframe (21,901 patients from February to June of 2020). The main study sample differed from this UNC target population in terms of age group, race, and sex characteristics, as seen in Table~\ref{tab:screennc}, and meta-analyses suggested that prevalence of COVID-19 infections differed between levels of these covariates in some populations \citep{pijls_demographic_2021, mackey_racial_2021}, supporting the covariates' use in standardization. Note that several racial classifications, including patient refused and unknown, were reclassified as `Other'. Second, standardization was made to the 2019 NC population over the age of 20 (7,873,971 persons) using covariate data from the American Community Survey \citep{us_census_bureau_american_2019}. The assumption of equal sampling probabilities may be less reasonable for this target population because not all NC residents are in the UNC Health Network and because there were some geographic areas where no patients in the study sample were from. There was no sample data in the main study for two covariate strata that existed in the UNC Health Network, so restriction was used for $\hat \pi_{SRG}$. Logistic regression models with main effects for sex, race, and age group were used to compute $\hat \pi_{SRGM}$; interaction effects were not included as the small number of positive test results could have led to model overfit. 

\begin{table}
	\caption{\label{tab:screennc}Demographic comparisons of the ScreenNC study sample, UNC Hospitals patient population, and North Carolina population aged 20+. Data on the NC population are from the 2019 American Community Survey \citep{us_census_bureau_american_2019}. Several racial classifications including Patient Refused and Unknown were reclassified as Other. Sample size is denoted by $n$. Some column totals do not sum to 100\% because of rounding.}
	\begin{tabular}{llcccccc} \hline
		&
		&
		\multicolumn{2}{c}{ScreenNC}& \multicolumn{2}{c}{UNC Hospitals}& \multicolumn{2}{c}{NC}  \\ 
		
		& & $n$ & \% & $n$ & \% & $n$ & \% \\ \hline
		
		& & 2,973 & 100 & 21,901 & 100 & 7,873,971 & 100 \\ \hline
		
		\multirow{2}{*}{Sex} & Female & 1,955 & 66 & 13,926 & 64 & 4,108,603 & 52\\
		
		& Male & 1,018 & 34 & 7,975 & 36 & 3,765,368 & 48 \\ \hline
		
		\multirow{4}{*}{Race} & Asian & 67 & 2 & 460 & 2 & 230,759 & 3 \\ 
		
		& Black or Af.-Am. & 395 & 13 & 5,109 & 23 & 1,640,311 & 21 \\
		
		& Other & 311 & 10 & 1,799 & 8  & 455,600 & 6 \\
		
		& White or Cauc. & 2,200 & 74 & 14,533 & 66 & 5,547,301 & 70 \\ \hline
		
		\multirow{7}{*}{Age} & 20-29 & 342 & 12 & 2060 & 9 & 1,400,918 & 18 \\ 
		
		& 30-39 & 599 & 20 & 2,763 & 13 & 1,344,647 & 17 \\
		
		& 40-49 & 518 & 18 & 3,382 & 15 & 1,351,156 & 17 \\
		
		& 50-59 & 602 & 20 & 4,200 & 19 & 1,360,357 & 17 \\
		
		& 60-69 & 489 & 17 & 4,548 & 21 & 1,228,123 & 16 \\		
		
		& 70-79 & 310 & 11 & 3,325 & 15 & 806,002 & 10 \\
		
		& 80+ & 77 & 3 & 1,623 & 7 & 382,768 & 5
		
	\end{tabular}
\end{table}

The sample proportion of positive tests was $\hat \rho =24/2973=0.81\%$. The sample false positive rate was $1 - \hat \sigma_p = 1.08\%$, so the data are, at first appearance, consistent with a population prevalence of 0\%. Indeed, the Rogan-Gladen seroprevalence estimate was $\hat \pi_{RG}=0$\% (95\% CI 0\%, 1.00\%). Likewise, the UNC target population had nonparametric and parametric standardized estimates of $\hat \pi_{SRG}=0$\% (0\%, 1.11\%) and $\hat \pi_{SRGM}=0$\% (0\%, 1.13\%), and the NC target population had corresponding estimates of 0\% (0\%, 1.10\%) and 0\% (0\%, 1.11\%). All estimates were truncated into $[0,1]$. The closeness of the standardized and unstandardized results may be due to the small number of positive test results and similarities between the sample and the target populations. Note that the limited violations of positivity and modest demographic differences (Table~\ref{tab:screennc}) make this application more similar to DGP 3 than DGP 4. Simulation results suggest the standardized estimators and corresponding CIs may perform well in settings similar to DGP 3, even if the true prevalence is low; e.g., see the lower right facets of Figure 2 and Appendix Figure~\ref{fig:dgp3_coverage}, where $\sigma_e=\sigma_p=.99$.

		\section{Discussion}
		\label{sec:discuss}
		
		Nonparametric and model-based standardized Rogan-Gladen estimators were examined, and their large-sample properties and consistent variance estimators were derived. While motivated by SARS-CoV-2 seroprevalence studies, the methods considered are also applicable to prevalence estimation of any binary variable for settings where validation data can be used to estimate the measurement instrument's sensitivity and specificity and covariate data can be used for standardization. Simulation studies demonstrated that both standardized Rogan-Gladen methods had low empirical bias and nominal CI coverage in the majority of practical settings. The empirical results in Section~\ref{sec:sims} highlight the tradeoffs inherent in choosing which method to use for a seroprevalence study. The parametric standardized estimator $\hat \pi_{SRGM}$ was empirically unbiased even when the number of strata and covariates, and with them the potential for random nonpositivity, increased. A drawback to $\hat \pi_{SRGM}$ is the need to correctly specify the form of a regression model. On the other hand, the nonparametric standardized estimator $\hat \pi_{SRG}$ does not require model specification and performed well in scenarios with lower amounts of selection bias and nonpositivity. As the number of strata and covariates grew, however, $\hat \pi_{SRG}$ was empirically biased and its corresponding 95\% CIs did not attain nominal coverage. 
		
		For practical use of either method, careful choice of covariates is necessary. Including additional covariates may make the assumption of equal probability of sampling within strata more reasonable. However, such inclusion also makes covariate-defined strata smaller and random nonpositivity more likely. An alternative strategy is to collapse smaller strata with few or no persons to create larger strata, which may make random nonpositivity less likely. However, if strata with sufficiently different sampling probabilities were collapsed together, then the assumption of equal probability of sampling within strata would be violated. 
		
	Throughout this paper it was assumed that the validation samples constitute random samples from the population strata of true positives and true negatives. Extensions could be considered which allow for possible ``spectrum bias'', which can occur if the $n_1$ individuals in the validation sample of true positives are not representative of the general population of seropositive individuals; for instance, a validation sample might be based on hospitalized patients who on average have more severe disease than the general population of seropositive individuals. Spectrum bias can lead to overestimates of sensitivity and underestimates of seroprevalence, but can be overcome using Bayesian mixture modeling of continuous antibody response data \citep{bottomley_quantifying_2021} or longitudinal modeling of antibody kinetic and epidemic data \citep{takahashi_sars-cov-2_2021}.
		
The limitations of Wald-type confidence intervals as they relate to parameters near their boundary values are mentioned in Sections~\ref{sec:methods_rg_estimators} and~\ref{sec:sims}. Alternative confidence intervals could be considered based on the bootstrap \citep{cai_exact_2022}, Bayesian posterior intervals \citep{gelman_bayesian_2020}, test inversion \citep{diciccio_confidence_2022}, or fiducial confidence distributions \citep{bayer_confidence_2022}. In particular, extensions of CIs designed to guarantee coverage at or above the nominal levels \citep{bayer_confidence_2022, lang_confidence_2014} could be developed to accommodate potential selection bias due to unknown sampling probabilities.
		
		Other topics for future research and broader issues in SARS-CoV-2 seroprevalence studies merit mention. While the approaches here estimate seroprevalence at a fixed point in time, seroprevalence is a dynamic parameter. For analysis of studies with lengthier data collection periods, extensions of the estimators in this paper could be considered which make additional assumptions (e.g., smoothness, monotonicity) about the longitudinal nature of seroprevalence. Another possible extension could consider variations in assay sensitivity, which may depend on a variety of factors such as the type of assay used; the recency of infection or vaccination of an individual; disease severity in infected individuals; the type and dose of vaccine for vaccinated individuals; and so forth. Where additional data are available related to these factors, then extensions of the standardized Rogan-Gladen estimators which incorporate these additional data could be developed. As an alternative to standardization, inverse probability of sampling weights \citep{lesko_generalizing_2017} or inverse odds of sampling weights \citep{westreich_transportability_2017} could be considered. standardization and weighting methods may possibly be combined to create a doubly robust Rogan-Gladen estimator.
			
			\bibliographystyle{apalike}
			\bibliography{refs3}
			
			\section*{Acknowledgments}
			
	We are grateful to Harm van Bakel, Juan Manuel Carreno, Frans Cuevas, Florian Krammer, and Viviana Simon for sharing the New York City seroprevalence data. We thank Dirk Dittmer and the ScreenNC research team for data access. We also thank Shaina Alexandria, Bryan Blette, and Kayla Kilpatrick for constructive suggestions. This research was supported by the NIH (Grant R01 AI085073), the UNC Chapel Hill Center for AIDS Research (Grants P30 AI050410 and R01 AI157758), and the NSF (Grant GRFP DGE-1650116). The content is solely the responsibility of the authors and does not represent the official views of the NIH.
			
			\section*{Code and data availability statement}
				R code based on the simulations in Section~\ref{sec:sims}, data supporting the Belgium and North Carolina seroprevalence study findings, and a Microsoft Excel spreadsheet that computes the estimators $\hat \pi_{SRG}$ and $\hat V_{\pi, SRG}$ are available at \url{https://github.com/samrosin/rgStandardized}. Data supporting the New York City seroprevalence study findings include both data available from \citet{stadlbauer_repeated_2021} and individual demographic data shared by the authors of that study.\vspace*{-8pt}
			
			\clearpage
			
			\begin{appendices}
							\setcounter{equation}{0}
				\renewcommand{\theequation}{A.\arabic{equation}}
				\section{Maximum likelihood estimation of $(\sigma_e, \sigma_p, \rho)$}
				\label{sec:mle}
								\renewcommand{\thesubsection}{A.\arabic{subsection}}
				\renewcommand{\theequation}{A.\arabic{equation}}
				
			Consider estimation of the parameter vector $(\sigma_e, \sigma_p, \rho)$. The likelihood is a product of three binomial distributions corresponding to the sensitivity, specificity, and main study datasets. Letting $T_1 = \sum_{i=1}^{n_1}X_i$, $T_2 = \sum_{i=n_1+1}^{n_1+n_2}X_i$, and $T_3=\sum_{i=n_1+n_2+1}^{n_1+n_2+n_3}X_i$, the log-likelihood is proportional to
				\begin{align*}
					&T_1\log \sigma_e + 
						(n_1 - T_1)\log(1-\sigma_e)+
						(n_2-T_2)\log\sigma_p \\
					&+ T_2\log(1 - \sigma_p)+ T_3\log\rho + (n_3 - T_3)\log(1- \rho) .
				\end{align*}
				Therefore the maximum likelihood estimator (MLE) of $(\sigma_e, \sigma_p, \rho)$ is $(\hat \sigma_e, \hat \sigma_p, \hat \rho)$ where $\hat \sigma_e = T_1 / n_1$, $\hat \sigma_p = (n_2 - T_2) / n_2$, and $\hat \rho = T_3 / n_3$. 
				
				\setcounter{equation}{0}
				\renewcommand{\theequation}{B.\arabic{equation}}
				\section{Proofs for Section 2} 
				\label{sec:proofs_rg}
				\renewcommand{\thesubsection}{B.\arabic{subsection}}
				\renewcommand{\theequation}{B.\arabic{equation}}
				
				\subsection{Proof of asymptotic normality}
				\label{sec:rg_can}
				Taylor expansion of $n^{-1}\sum\psi(X_i;\delta_i,\hat \theta)$ around the true parameter $\theta$ yields \[ 
				0=n^{-1}\sum\psi(X_i;\delta_i,\hat \theta) 
				=
				n^{-1}\sum\psi(X_i;\delta_i,\theta)
				+n^{-1}\sum\dot\psi(X_i;\delta_i,\theta)(\hat\theta - \theta)
				+R 
				\]
				where $\dot \psi(X_i; \delta_i,\theta) = \partial \psi(X_i; \delta_i,\theta)/\partial \theta^T$ and $R$ is a remainder term. Rearranging and multiplying by $\sqrt{n}$ yields
				\begin{equation}
					\sqrt{n}(\hat \theta - \theta) =
					\left\{
					n^{-1}\sum -\dot\psi(X_i; \delta_i, \theta) 
					\right\}^{-1}
					\sqrt{n}
					\left\{
					n^{-1}\sum \psi(X_i;\delta_i,\theta)
					\right\}
					+\sqrt{n} R^*
					\label{eq:taylor}
				\end{equation}
				where $R^*$ is a new remainder defined below. It is shown below that $n^{-1}\sum -\dot\psi(X_i; \delta_i, \theta) \to_p \mathbb{A}(\theta)$, $\sqrt{n} \{n^{-1}\sum \psi(X_i;\delta_i,\theta)\} \to_{d} \mathcal{N}\{0,\mathbb B(\theta)\}$, and $\sqrt{n}R^*\to_{p}0$, where $\mathbb{A}(\theta)$ and $\mathbb{B}(\theta)$ are defined below. Therefore, by Slutsky's theorem, $
				\sqrt{n}(\hat \theta - \theta) \to_{d} \mathcal{N}\left\{0, \mathbb A(\theta)^{-1}
				\mathbb B(\theta) 
				\mathbb A(\theta)^{-T}\right\} 
				$. 
				
				First, define \begin{align*}
					\mathbb{A}_n(X,\delta,\theta)&=n^{-1}\sum -\dot\psi(X_i;\delta_i,\theta) \\
					&=n^{-1}\sum
					\left[
					\begin{array}{cccc}
						I(\delta_i=1) &0 &0 &0\\
						0 &I(\delta_i=2)&0 &0\\
						0 &0 &I(\delta_i=3) &0\\
						\pi & -1+\pi &-1 &\sigma_{e} + \sigma_{p} -1 
					\end{array}
					\right]
				\end{align*}
				and let
				\begin{align*}  
					\mathbb{A}(\theta)=
					\left[
					\begin{array}{cccc}
						c_1 &0 &0 &0\\
						0 &c_2&0 &0\\
						0 &0 &c_3 &0\\
						\pi & -1+\pi &-1 &\sigma_{e} + \sigma_{p} -1 
					\end{array}
					\right]. 
				\end{align*}
				As $n \to \infty$, $\mathbb A_{n}(X,\delta,\theta)\to \mathbb A(\theta)$ by the assumption that $n^{-1}\sum I(\delta_i = j) = n_j/n\to c_j\in(0, 1)$ for $j \in \{1,2,3\}$. 
				
				Second, for brevity, let $\psi_e$ denote $\psi_e(X_i;\delta_i,\theta_i)$, and similarly for $\psi_p, \psi_\rho$, and $\psi_\pi$. Define 
				\begin{align*}
					\mathbb{B}_n(X,\delta,\theta) &= n^{-1}\sum \mathbb E\left\{\psi(X_i;\delta_i,\theta)\psi(X_i;\delta_i,\theta)^T\right\}\\
					&=n^{-1}\sum \mathbb E
					\left[
					\begin{array}{cccc}
						\psi_e^2 &0 &0 &\psi_e \psi_{\pi}\\
						0 &\psi_p^2 &0 &\psi_p \psi_{\pi}\\
						0 &0 & \psi_{\rho}^2 &\psi_{\rho} \psi_{\pi}\\
						\psi_{\pi} \psi_e &\psi_{\pi} \psi_p &\psi_{\pi} \psi_{\rho} & \psi_{\pi}^2
					\end{array}
					\right]
				\end{align*} and let
				\[
				\mathbb{B}(\theta)=	\left[
				\begin{array}{cccc}
					c_1\sigma_e (1-\sigma_e) &0 &0 &0\\
					0 &c_2\sigma_p (1-\sigma_p)  &0 &0\\
					0 &0 &c_3{\rho} (1-{\rho})  &0\\
					0 &0 &0  & 0
				\end{array}
				\right].
				\]
				Note that $\mathbb{B}(\theta)=\lim_{n\to\infty}\mathbb{B}_n(X,\delta,\theta)$
				as $
				\mathbb E( \psi_e^2 )= \mathbb E \{I(\delta_i=1) (X_i-\sigma_{e})^2 \}= I(\delta_i = 1)\sigma_{e} (1-\sigma_{e})
				$ and likewise for $\mathbb E(\psi_p^2)$ and $\mathbb E(\psi_\rho^2)$. It follows that $
				\sqrt{n}\left\{n^{-1}\sum \psi(X_i;\delta_i,\theta)\right\}\to_{d}\mathcal{N}\{0,\mathbb B(\theta)\}
				$
				by the Lindeberg-Feller CLT. In particular, let $s_n^2=\sum\{\operatorname{Var}(\psi_e)+\operatorname{Var}(\psi_p)+\operatorname{Var}(\psi_{\rho})+\operatorname{Var}(\psi_{\pi})\}=n_1\sigma_e(1-\sigma_e) + n_2\sigma_p(1-\sigma_p)+n_3\rho(1-\rho)$ and let $\lVert\cdot\rVert$ denote the Euclidean norm. Note that $\max_{i}\lVert\psi(X_i;\delta_i,\theta)\rVert\leq\sqrt{2}$
				as the maximum magnitude of each element of $\psi(X_i; \delta_i, \theta)$ is 1, and for a given observation $X_i$ it is always true that two of the indicators $\delta_1, \delta_2, \delta_3$ equal zero and the third indicator equals one. Therefore for all $\epsilon>0$,	
				\[	\lim_{n\to\infty} s_n^{-2}\sum \mathbb E\left\{
				\lVert\psi(X_i;\delta_i,\theta)\rVert^2I(\lVert\psi(X_i;\delta_i,\theta)\rVert
				\geq 
				\epsilon s_n)
				\right\} 
				\leq 
				\lim_{n\to\infty} 2s_n^{-2}\sum P(\sqrt{2}\geq \epsilon s_n)
				=
				0,
				\] 
				implying the Lindeberg condition holds.
				
				Third, it remains to prove $\sqrt{n}R^*\to_{p}0$. The outline of the proof of Boos and Stefanski (2013) Theorem 7.2 can be followed, but their assumption of identically distributed data must be removed. Consider the second-order Taylor series expansion of the $j$th element of the vector $n^{-1}\sum \psi(X_i;\delta_i,\hat\theta)$, denoted $n^{-1}\sum \psi_j(X_i;\delta_i,\hat\theta)$, around the true value $\theta$:
				\begin{align*}
					0=n^{-1}\sum \psi_j(X_i;\delta_i,\hat\theta)
					&=n^{-1}\sum \psi_j(X_i;\delta_i,\theta) 
					+ n^{-1}\sum \dot\psi_j(X_i;\delta_i,\theta)(\hat \theta - \theta)\\
					&+\frac{1}{2} (\hat \theta - \theta)^Tn^{-1}
					\sum \ddot\psi_j(X_i;\delta_i,\widetilde\theta_j)(\hat \theta - \theta)
				\end{align*}
				where $\widetilde \theta_1,\dots,\widetilde\theta_4$ are on the line segment joining $\hat \theta$ and $\theta$ and $\ddot\psi_j(X_i;\delta_i,\theta_j)$ is a $4\times 4$ matrix with entry $(j, k)$ equal to $\partial^2 \psi_j(X_i; \delta_i,\theta_j)/\partial \theta_j\partial\theta_k$ for $j, k \in \{1, 2,3,4\}$. Writing these 4 equations in matrix notation yields 
				\begin{equation}
					0=n^{-1}\sum \psi(X_i;\delta_i,\theta)+\widetilde{R}(\hat \theta - \theta)
					\label{eq:matrix}
				\end{equation}
				where $\widetilde{R} = \left\{n^{-1}\sum \dot\psi(X_i;\delta_i,\theta)+(1/2)\widetilde Q\right\}$ and \[
				\widetilde{Q} = \left[ \begin{array}{cccc}
					0 & 0 & 0 & 0 \\
					0 & 0 & 0 & 0 \\
					0 & 0 & 0 & 0 \\
					\sigma_e - \hat \sigma_e & \sigma_p - \hat \sigma_p & 0 & 0 
				\end{array}
				\right]\] 
				is the $4\times4$ matrix with $j$th row given by $(\hat \theta - \theta)^Tn^{-1}\sum \ddot\psi_j(X_i;\delta_i,\widetilde\theta_j)$. Note that $\widetilde Q \to_{p} 0_{4\times 4}$ by the Weak Law of Large Numbers, where in general $0_{r \times c}$ denotes an $r \times c$ matrix of zeros. Also note that $n^{-1}\sum \dot\psi(X_i;\delta_i,\theta) \to_p -\mathbb{A}(\theta)$, where $-\mathbb{A}(\theta)$ is nonsingular under the assumption that $\sigma_e > 1 - \sigma_p$. It follows that as $n\to\infty$, $\widetilde{R}$ is invertible with probability one. On the set $S_n$ where $\widetilde{R}$ is invertible, \eqref{eq:matrix} can be rearranged to yield
				\begin{equation}
					\hat \theta - \theta = 
					\left(-\widetilde{R}\right)
					^{-1} 
					\left\{n^{-1}\sum \psi(X_i;\delta_i, \theta)\right\}.
					\label{eq:penult}
				\end{equation}
				Define \[\widetilde{R}^* = \frac{1}{1+g}\left\{n^{-1}\sum -\dot\psi(X_i;\delta_i,\theta)\right\}^{-1} (1/2)\widetilde{Q}\left\{n^{-1}\sum -\dot\psi(X_i;\delta_i,\theta)\right\}^{-1} \] 	where $g=\mathrm{Tr}\left[-(1/2)\widetilde Q\left\{n^{-1}\sum -\dot\psi(X_i;\delta_i,\theta)\right\}^{-1}\right]$, and note that $\widetilde{R}^*\to_p 0_{4\times4}$ by Slutsky's theorem because $\widetilde{Q}\to_p 0_{4 \times 4}$.  Since $(1/2)\widetilde{Q}$ has rank one, $\left(-\widetilde{R}\right)^{-1}=\left\{n^{-1}\sum 
				-	\dot\psi(X_i;\delta_i,\theta)\right\}^{-1} + \widetilde{R}^*$ by an application of the Sherman-Morrison-Woodbury formula (Miller, 1981). Thus, substituting this expression for $(-\widetilde{R})^{-1}$ into \eqref{eq:penult} and multiplying by $\sqrt{n}$ yields \eqref{eq:taylor}, where
				$R^*=\widetilde{R}^* \left\{n^{-1}\sum \psi(X_i;\delta_i, \theta)\right\}$. Therefore, because $\lim_{n\to\infty}\Pr(S_n)=1$,  $\widetilde{R}^*\to_p 0_{4\times4}$, and $\sqrt{n}\left\{n^{-1}\sum \psi(X_i;\delta_i, \theta)\right\} \to_d \mathcal{N}(0, \mathbb{B}(\theta))$, 
				by Slutsky's theorem $\sqrt{n}R^* \to_p 0$.
				
				\subsection{Computation of asymptotic variance}
				
				Since $\mathbb A(\theta)$ is lower triangular, it follows that
				\[
				\mathbb A(\theta)^{-1}=
				\left[
				\begin{array}{cccc}
					1/c_1 &0 &0 &0\\
					0 &1/c_2&0 &0\\
					0 &0 &1/c_3&0\\
					- \dfrac{\pi}{c_1 (\sigma_{e} + \sigma_{p}-1)}  
					& \dfrac{1-\pi}{c_2(\sigma_{e} + \sigma_{p}-1)} 
					& \dfrac{1}{c_3(\sigma_{e} + \sigma_{p}-1)} 
					&(\sigma_{e} + \sigma_p -1)^{-1} 
				\end{array}
				\right],
				\]
				and therefore
				$\mathbb A(\theta)^{-1}
				\mathbb B(\theta) 
				\mathbb A(\theta)^{-T}$ equals
				\[
				\left[
				\begin{array}{cccc}
					\sigma_e (1-\sigma_e)/c_1 &0 &0 &*\\
					0 & \sigma_p (1-\sigma_p)/c_2 &0 &*\\
					0 &0 &{\rho} (1-{\rho})/c_3 &*\\
					* & *
					&  *
					& V_{\pi,RG}
				\end{array}
				\right],
				\]
				where * denotes quantities not expressed explicitly and
				\[
				V_{\pi,RG} =
				\left\{
				\frac{\pi^2 \sigma_e (1-\sigma_e)}
				{c_1}
				+
				\frac{ (1-\pi)^2 \sigma_p (1-\sigma_p)}{c_2} 
				+ 
				\frac{\rho(1-{\rho})}{c_3}
				\right\}
				(\sigma_e + \sigma_p -1)^{-2}.
				\]  
				
				\section{Proofs for Section 3.2}
				\label{sec:proofs_srg}
				
				\setcounter{subsection}{0}
				
				\renewcommand{\thesubsection}{C.\arabic{subsection}}
				
				\subsection{Proof of asymptotic normality}
				\label{sec:srg_can}
				
				The Taylor expansion of $n^{-1}\sum\psi(X_i,Z_i;\delta_i,\hat\theta_s)$ around the true parameter $\theta_s$ yields 
				\[
				\sqrt{n}(\hat \theta_s - \theta_s) =
				\left\{
				n^{-1}\sum -\dot\psi(X_i, Z_i;\delta_i, \theta_s) 
				\right\}^{-1}
				\sqrt{n}\left\{
				n^{-1}\sum \psi(X_i,Z_i;\delta_i,\theta_s)
				\right\}
				+
				\sqrt{n}R^*,
				\]
				which is similar in form to \eqref{eq:taylor}, except here the estimating equations are dependent on covariates $Z$. The remainder $R^*$ here is distinct from that in Appendix~\ref{sec:rg_can}, and in general symbols may be reused and notation may not hold the same meaning across appendices. Below it is established, using an analogous approach to that of Appendix~\ref{sec:rg_can}, that $n^{-1}\sum -\dot\psi(X_i, Z_i; \delta_i, \theta_s) \to_p \mathbb{A}(\theta_s)$, $\sqrt{n} \{n^{-1}\sum \psi(X_i, Z_i;\delta_i,\theta_s)\} \to_{d} \mathcal{N}\{0,\mathbb B(\theta_s)\}$, and $\sqrt{n}R^*\to_{p}0$. Therefore, by Slutsky's theorem,
				$\sqrt{n}(\hat \theta_s - \theta_s) \to_{d} \mathcal{N}\{
				0,\mathbb A(\theta_s)^{-1}\mathbb B(\theta_s)\mathbb A(\theta_s)^{-T}
				\}$. 
				
				First, define 
				\begin{align*}
					\mathbb{A}_n(X,Z,\delta,\theta_s)=n^{-1}\sum \left\{
					-\dot\psi(X_i,Z_i;\delta_i,\theta_s) 
					\right\}
					=
					\left[\begin{array}{cc}
						A' & 0_{(k+2) \times 2} \\
						C & D	
					\end{array}\right]
				\end{align*} as a block matrix where $A' =  \operatorname{diag}\left(n_1/n,n_2/n,n_{z_1}/n,\dots,n_{z_k}/n\right)$ is $(k+2)\times (k+2)$, \[
				C=\left[\begin{array}{ccccc}
					0 & 0 & -\gamma_1 & \dots & -\gamma_k \\ 
					\pi & -1+\pi& 0 & \dots & 0 \end{array}\right]
				\]
				is $2 \times (k+2)$, and
				\[
				D = \begin{bmatrix} 1 & 0 \\ -1 & \sigma_e+\sigma_p -1 \end{bmatrix}.
				\] Let	\[ 
				\mathbb A(\theta_s)=\left[\begin{array}{cc}
					A & 0_{(k+2)\times2} \\
					C & D	
				\end{array}\right] 
				\] where $A=\operatorname{diag}\{c_1,c_2,c_3s_{1},\dots,c_3s_{k} \}$, and note that $\mathbb{A}(\theta_s) = \lim_{n\to\infty}\mathbb{A}_n(X,Z,\delta,\theta_s)$ since $n_j/n\to c_j$ for $j\in\{1, 2, 3\}$. Note that $s_1, \dots, s_k$ are all nonzero due to positivity, without which $\hat \pi_{SRG}$ is undefined. 
				
				Second, define 
				\begin{align*}
					\mathbb B_n(X,Z,\delta,\theta_s)=n^{-1}\sum \mathbb E\{
					\psi(X_i,Z_i;\delta_i,\theta_s)\psi(X_i,Z_i;\delta_i,\theta_s)^T
					\}
					=
					\begin{bmatrix}E' & 0_{(k+2)\times2} \\ 0_{2\times(k+2)} & 0_{2\times2} \end{bmatrix}, 
				\end{align*} where 
				\[
				E'=\operatorname{diag}\{n_1\sigma_e (1-\sigma_e)/n,n_2 \sigma_p (1-\sigma_p)/n,n_{z_1} \rho_1 (1-\rho_1)/n,\dots,n_{z_k}\rho_k(1-\rho_k)/n\},
				\]
				since $\mathbb{E}(\psi_{\rho_j}^2)=\mathbb{E}\{I(Z_i=z_j, \delta_i = 3)(X_i - \rho_j)^2\}=I(Z_i = z_j, \delta_i = 3)\rho_j(1-\rho_j)$.
				Define $\mathbb B(\theta_s)$ to have the same form as $\mathbb B_n(X,Z,\delta,\theta_s)$ except $E'$ is replaced by 
				\[
				E=\operatorname{diag}\{c_1\sigma_e (1-\sigma_e),c_2 \sigma_p (1-\sigma_p),c_3 s_{1} \rho_1 (1-\rho_1),\dots,c_3 s_{k}\rho_k(1-\rho_k)\}.
				\] Since $\mathbb{B}(\theta_s)=\lim_{n\to\infty}\mathbb{B}_n(X,Z,\delta,\theta_s)$, $ \sqrt{n}
				\left\{
				n^{-1}\sum \psi(X_i, Z_i; \delta_i,\theta_s)
				\right\}
				\to_{d}
				\mathcal{N}\{
				0,\mathbb B(\theta_s)
				\}
				$ by the Lindeberg-Feller CLT. 
				In particular, let \begin{align*}
					v_n^2&=\sum\{\operatorname{Var}(\psi_e)+\operatorname{Var}(\psi_p)+\operatorname{Var}(\psi_{\rho_1})+\dots+\operatorname{Var}(\psi_{\rho_k})+\operatorname{Var}(\psi_{\rho})+\operatorname{Var}(\psi_{\pi})\}\\
					&=n_1\sigma_e(1-\sigma_e) + n_2\sigma_p(1-\sigma_p)+\sum_{j=1}^kn_{z_j} \rho_j(1-\rho_j),
				\end{align*} and note that $\max_i \lVert\psi(X_i,Z_i;\delta_i,\theta_s)\rVert\leq\sqrt{3}$ because 
				at most one of $\psi_e, \psi_p, \psi_{\rho_1}, \dots, \psi_{\rho_k}$ is nonzero and each element of $\psi(X_i, Z_i; \delta_i, \theta_s)$ has a maximum value of one. Therefore for all $\epsilon>0$,	
				\[	\lim_{n\to\infty} v_n^{-2}\sum \mathbb E\left\{
				\lVert\psi(X_i;\delta_i,\theta)\rVert^2I(\lVert\psi(X_i;\delta_i,\theta)\rVert
				\geq 
				\epsilon v_n)
				\right\} 
				\leq 
				\lim_{n\to\infty} 3v_n^{-2}\sum P(\sqrt{3}\geq \epsilon v_n)
				=
				0,
				\]  
				implying the Lindeberg condition holds.
				
				Third, let $\widetilde Q_s$ be analogous to $\widetilde Q$ from Appendix~\ref{sec:rg_can}. Denote the $j$th entry of $\theta_s$ as $\theta_{s_j}$ and note $\lvert \partial^2\psi_j(X_i,Z_i;\delta_i,\theta_s) / \partial \theta_{s_j}\theta_{s_l} \rvert \leq 1$ for all $j, l \in\{1,2,\dots,k+4\}$. Thus there exists a function $g(X_i,Z_i,\delta_i)$ such that, for each $\theta_s^*$ in a neighborhood of $\theta_s$, $\left\lvert \partial^2 \psi_j(X_i,Z_i;\delta_i,\theta_s) / \partial \theta^*_{s_j} \partial \theta^*_{s_l}\right\rvert\leq g(X_i,Z_i,\delta_i)$ for all $X_i, Z_i,\delta_i$ where $\int g(X_i,Z_i,\delta_i) dF(X_i,Z_i,\delta_i)<\infty$; e.g., let $g(X_i,Z_i,\delta_i)=2$. Therefore, each entry in $\widetilde Q_s$ is bounded by $\lVert\hat \theta_s - \theta_s\rVert n^{-1}\sum g(X_i,\delta_i)=o_p(1)$, so $\widetilde Q_s \to_{p} 0_{(k+4)\times(k+4)}$. This fact can be used to prove $\sqrt{n}R^*\to_{p}0$ with the same technique as in Appendix~\ref{sec:rg_can}, so we omit the rest of the proof. 
				
				\subsection{Computation of asymptotic variance}
				
				Note that $\mathbb A(\theta_s)$ is block lower triangular, and thus
				\[
				\mathbb A(\theta_s)^{-1}=\begin{bmatrix} A^{-1} & 0_{(k+2)\times2} \\ -D^{-1}CA^{-1} & D^{-1} \end{bmatrix}
				\] and therefore
				\begin{align*}
					\mathbb A(\theta_s)^{-1}\mathbb B(\theta_s)\mathbb A(\theta_s)^{-T} 
					&= \begin{bmatrix} A^{-1}EA^{-1} & -A^{-1}EA^{-1}C^TD^{-T} \\ -D^{-1}CA^{-1}EA^{-1} & D^{-1}CA^{-1}EA^{-1}C^TD^{-T} \end{bmatrix}.
				\end{align*} The bottom right submatrix of $	\mathbb A(\theta_s)^{-1}\mathbb B(\theta_s)\mathbb A(\theta_s)^{-T} $, and specifically the bottom right element of that submatrix, is of primary interest. Since $A$ and $E$ are both diagonal,
				\[
				A^{-1}EA^{-1} = \operatorname{diag}\{c_1^{-1}\sigma_e(1-\sigma_e),c_2^{-1}\sigma_p(1-\sigma_p), 
				c_3^{-1}s_{1}^{-1}\rho_1(1-\rho_1),\dots,c_3^{-1}s_{k}^{-1}\rho_k(1-\rho_k)\}.
				\]
				Next, note
				\[
				D^{-1}C=\begin{bmatrix} 0 & 0 & -\gamma_1 & \dots & -\gamma_k \\ \dfrac{\pi}{\sigma_e+\sigma_p-1} & \dfrac{-1+\pi}{\sigma_e+\sigma_p-1} & \dfrac{-\gamma_1}{\sigma_e+\sigma_p-1} & \dots & \dfrac{-\gamma_k}{\sigma_e+\sigma_p-1} \end{bmatrix}.
				\] Therefore, $	D^{-1}CA^{-1}EA^{-1}$ equals
				\begin{align*}
					\begin{bmatrix}
						0 & 0 & -\gamma_1c_{3}^{-1} s_{1}^{-1}\rho_1(1-\rho_1) & \dots & -\gamma_kc_{3}^{-1} s_{k}^{-1}\rho_1(1-\rho_1) ) \\	
						\dfrac{\pi\sigma_e(1-\sigma_e)}{c_1(\sigma_e+\sigma_p-1)} & \dfrac{(-1+\pi)\sigma_p(1-\sigma_p)}{c_2(\sigma_e+\sigma_p-1)} & -\dfrac{\gamma_1\rho_1(1-\rho_1)}{c_3s_{1}(\sigma_e+\sigma_p-1)} & \dots & -\dfrac{\gamma_k\rho_k(1-\rho_k)}{c_3s_{k}(\sigma_e+\sigma_p-1)} \end{bmatrix}
				\end{align*}
				and thus \[
				D^{-1}CA^{-1}EA^{-1}C^TD^{-T}=\left[\begin{array}{cc} * & \ * \\ * & V_{\pi,SRG} \end{array} \right]
				\]
				where * denotes quantities not expressed explicitly and 
				\[
				V_{\pi,SRG}= 
				\left\{
				\frac{\pi^2 \sigma_e (1-\sigma_e)}
				{c_1}
				+
				\frac{ (1-\pi)^2 \sigma_p (1-\sigma_p)}{c_2} 
				+ 
				\sum_{j=1}^k\frac{\gamma_j^2\rho_j(1-{\rho_j})}{c_3s_{j}}
				\right\}
				(\sigma_e + \sigma_p -1)^{-2}.
				\]

				\section{Proof for Section 3.3}
				\label{sec:proofs_srgm}
				\setcounter{subsection}{0}
				\renewcommand{\thesubsection}{D.\arabic{subsection}}
				
				The proof that $\theta_m$ is asymptotically normal is given for the case of logistic regression, but it extends to any link function $g(\cdot)$ appropriate for binary regression. Recall that the $(p+4)$-vector of estimating equations is \[\sum \psi(X_i,Z_i; \delta_i,\theta_m)=\left(\sum \psi_e, \sum \psi_p, \sum \psi_\beta,  \psi_\rho,\psi_\pi\right)^T=0.
				\]
				In the above vector $\sum \psi_e, \sum \psi_p$, and $\psi_\pi$ are identical to the equations used in Appendices~\ref{sec:proofs_rg} and \ref{sec:proofs_srg}; $\psi_\rho
				=
				\sum_{j=1}^k  \operatorname{logit}^{-1}  \{
				\beta h(Z_j)
				\}\gamma_j-\rho$; and $\sum \psi_\beta$ is a $p$-vector with $j$th element $\sum \psi_{\beta_j}=\sum I(\delta_i=3)\left[
				X_i-\operatorname{logit}^{-1}
				\{\beta h(Z_i)\}
				\right]h_j(Z_i)$. The Taylor expansion of $n^{-1}\sum\psi(X_i,Z_i;\delta_i,\hat\theta_m)$ around the true parameter $\theta_m$ yields 
				\[
				\sqrt{n}(\hat \theta_m - \theta_m) =
				\left\{
				n^{-1}\sum -\dot\psi(X_i, Z_i;\delta_i, \theta_m) \right\}^{-1}
				\sqrt{n}\left\{
				n^{-1}\sum	 \psi(X_i,Z_i;\delta_i,\theta_m)
				\right\}
				+\sqrt{n}R^*.
				\] 
				
				The rest of the proof is similar to Appendices~\ref{sec:rg_can} and \ref{sec:srg_can}. Namely, first define \[
				\mathbb{A}_n(X,Z,\delta,\theta_m)=n^{-1}\sum 
				\left\{
				-\dot\psi(X_i,Z_i;\delta_i,\theta_m) 
				\right\}
				=\left[\begin{array}{ccc}
					A' & 0_{2\times p} & 0_{2\times2} \\
					0_{p\times2} & B' & 0_{p\times 2} \\ 
					C & D & E	
				\end{array}\right]
				\]
				as a block matrix where $A'=\operatorname{diag}(n_1/n,n_2/n)$, $B'$ is $p\times p$ with entry $(j, k)$ equal to $n^{-1}\sum I(\delta_i=3) h_j(Z_i)h_k(Z_i)\exp\{
				\beta h(Z_i)\}/ \left[
				1+\exp\{\beta h(Z_i)\}
				\right]^2$, \[C = \begin{bmatrix} 0 & 0 \\ \pi & -1 + \pi \end{bmatrix},
				\] 
				\[
				D=\left[\begin{array}{ccccc}
					-\sum_{j=1}^k h_1(Z_j)\dfrac{
						\exp\{
						\beta h(Z_i)\}
					}{\left[
						1+\exp\{\beta h(Z_i)\}
						\right]^2
					}\gamma_j & \dots &  -\sum_{j=1}^k h_p(Z_j)\dfrac{
						\exp\{
						\beta h(Z_i)\}
					}{\left[
						1+\exp\{\beta h(Z_i)\}
						\right]^2
					}\gamma_j  \\ 
					0 & \dots & 0 \end{array}\right]
				\]
				is $2 \times p$, and 
				\[E = \begin{bmatrix} 1 & 0 \\ -1 & \sigma_e+\sigma_p -1 \end{bmatrix}.\] Let $\mathbb A(\theta_m)$ have the same form as $\mathbb{A}_n(X,Z,\delta,\theta_m)$, replacing $A'$ with $A = \operatorname{diag}(c_1, c_2)$ and $B'$ with $B$, with entry $(j, k)$ of $B$ equal to $c_3\mathbb E\left( h_j(Z)h_k(Z)
				\exp\{
				\beta h(Z_i)\} / 
				\left[
				1+\exp\{\beta h(Z_i)\}
				\right]^2
				\right)$. Note that $\mathbb A_n(X, Z, \delta, \theta_m)\to_p\mathbb{A}(\theta_m)$ by the Weak Law of Large Numbers.
				
				Second, define 
				\[
				\mathbb B_n(X,Z,\delta,\theta_m)=n^{-1}\sum \mathbb E(\psi(X_i,Z_i;\delta_i,\theta_m)\psi(X_i,Z_i;\delta_i,\theta_m)^T)=\left[\begin{array}{ccc}
					F' & 0_{2\times p} & 0_{2 \times 2} \\
					0_{p \times 2} & G' & 0_{p \times 2}\\ 
					0_{2\times 2} & 0_{2 \times p} & 0_{2 \times 2}	
				\end{array}\right]
				\]
				where $F'=\operatorname{diag}\left\{n_1\sigma_e(1-\sigma_e)/n,n_2\sigma_p(1-\sigma_p)/n\right\}$ and $G'$ is a $p\times p$ matrix with entry $(j, k)$ equal to $(n_3/n)	\mathbb E\left(h_j(Z)h_k(Z)
				\left[X-\operatorname{logit}^{-1}\{\beta h(Z)\}\right]^2\right)$. Let $\mathbb B(\theta_m)$ be of the same form as $\mathbb B_n(X,Z,\delta,\theta_m)$ except with $F$ and $G$ replacing $F'$ and $G'$, where $F=\operatorname{diag}\{c_1\sigma_e(1-\sigma_e),c_2\sigma_p(1-\sigma_p)\}$ and $G$ is identical to $G'$ except with $c_3$ replacing $n_3/n$ in each element. Noting that $\mathbb{B}(\theta_m)=\lim_{n\to\infty}\mathbb{B}_n(X,Z,\delta,\theta_m)$, it follows that $
				\sqrt{n}\left\{n^{-1}\sum \psi(X_i,Z_i;\delta_i,\theta_m)\right\}\to_{d}\mathcal{N}\left\{
				0,\mathbb B(\theta_m)
				\right\}
				$ by the Lindeberg-Feller CLT. Specifically, note that the range of the inverse logit function is $[0, 1]$ and that all elements of $h(Z)$, the user-specified function of the covariates, must be finite. Then the Lindeberg condition holds by the same logic as in Appendices~\ref{sec:rg_can} and \ref{sec:srg_can}. 
				
				Finally, $\sqrt{n}R^*\to_{p}0$ can be proved as in Appendix~\ref{sec:srg_can}. Letting $\theta_{m_j}$ denote the $j$th entry of $\theta_m$, the proof follows from the fact that $\lvert \partial^2\psi_j(X_i,Z_i;\delta_i,\theta_m) / \partial \theta_{m_j}\partial\theta_{m_k}\rvert < \infty$ for all $j,k \in\{1,2,\dots,p+4\}$. Therefore, by Slutsky's Theorem
				\[
				\sqrt{n}(\hat \theta_m - \theta_m) \to_{d} \mathcal{N}\left\{
				0,\mathbb A(\theta_m)^{-1}\mathbb B(\theta_m)\mathbb A(\theta_m)^{-T}
				\right\} \ \text{and}  \ 
				\sqrt{n}(\hat \pi - \pi) \to_d \mathcal{N}(0, V_{\pi, SRGM}) 
				\]
				where the bottom right element of $\mathbb A(\theta_m)^{-1}\mathbb B(\theta_m)\mathbb A(\theta_m)^{-T}$ is denoted as $V_{\pi, SRGM}$. The asymptotic variance of $\hat \theta_m$ can be consistently estimated by the empirical sandwich variance estimator $\mathbb A(\hat \theta_m)^{-1} \mathbb B(\hat \theta_m) \mathbb A(\hat \theta_m)^{-T}$; the bottom right element of the sandwich estimator is $\hat V_{\pi, SRGM}$.
				\clearpage
				
				\setcounter{figure}{0}    
				\renewcommand\thefigure{A.\arabic{figure}}    
				\renewcommand{\theHfigure}{A\arabic{figure}}
				
				\section*{Appendix Figures}
				\setcounter{subsection}{0}
				\renewcommand{\thesubsection}{E.\arabic{subsection}}
			\begin{figure}[ht!]
				\centering
				\includegraphics[width=1\textwidth]{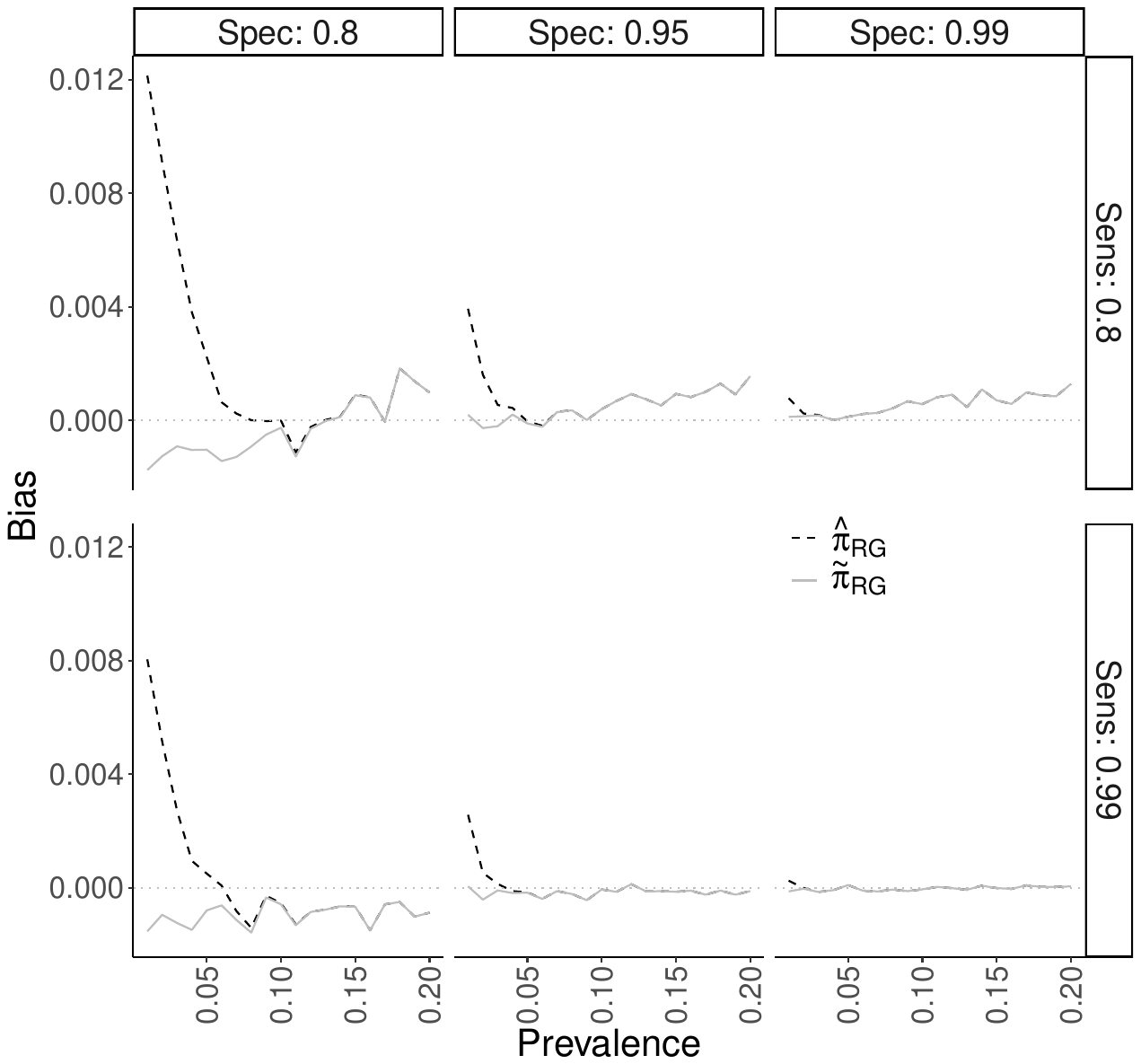}
				\caption{
					Empirical bias of the Rogan-Gladen ($\hat \pi_{RG}$) estimator and the non-truncated Rogan-Gladen estimator ($\tilde \pi_{RG}$) from simulation study for DGP 1, described in Section 4.1 of the main text. The six facets correspond to a given combination of sensitivity $\sigma_e$ (`Sens') and specificity $\sigma_p$ (`Spec'). 10,000 simulations were conducted for this scenario.}
				\label{fig:dgp1_bias}
			\end{figure}
			
			\begin{figure}
				\centering
				\includegraphics[width=1\textwidth]{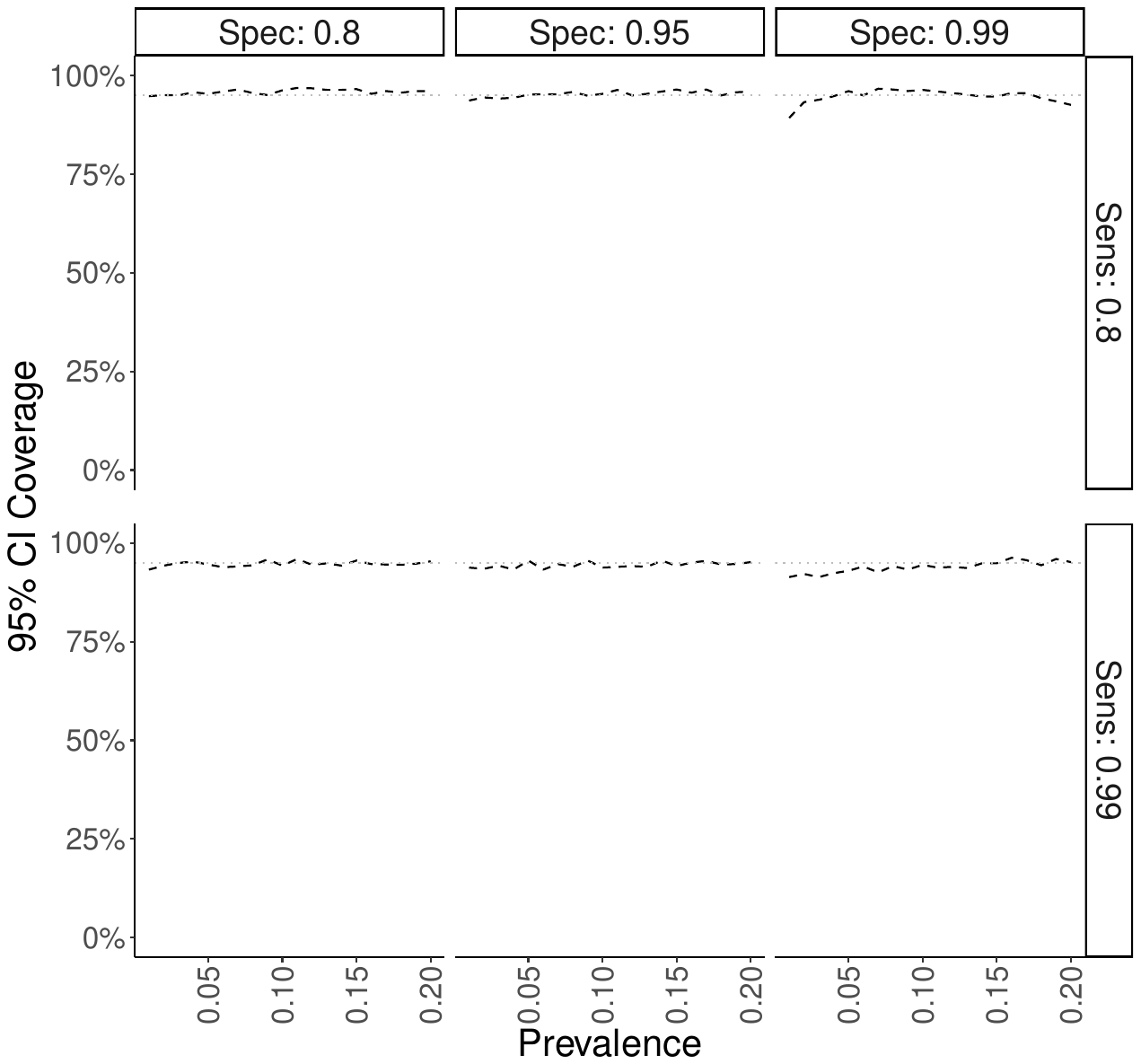}
				\caption{Confidence interval coverage of the Rogan-Gladen ($\hat \pi_{RG}$) estimator from simulation study for DGP 1, described in Section 4.1 of the main text. 10,000 simulations were conducted for this scenario.}
				\label{fig:dgp1_coverage}
			\end{figure}

			\begin{figure}
				\centering
				\includegraphics[width=1\textwidth]{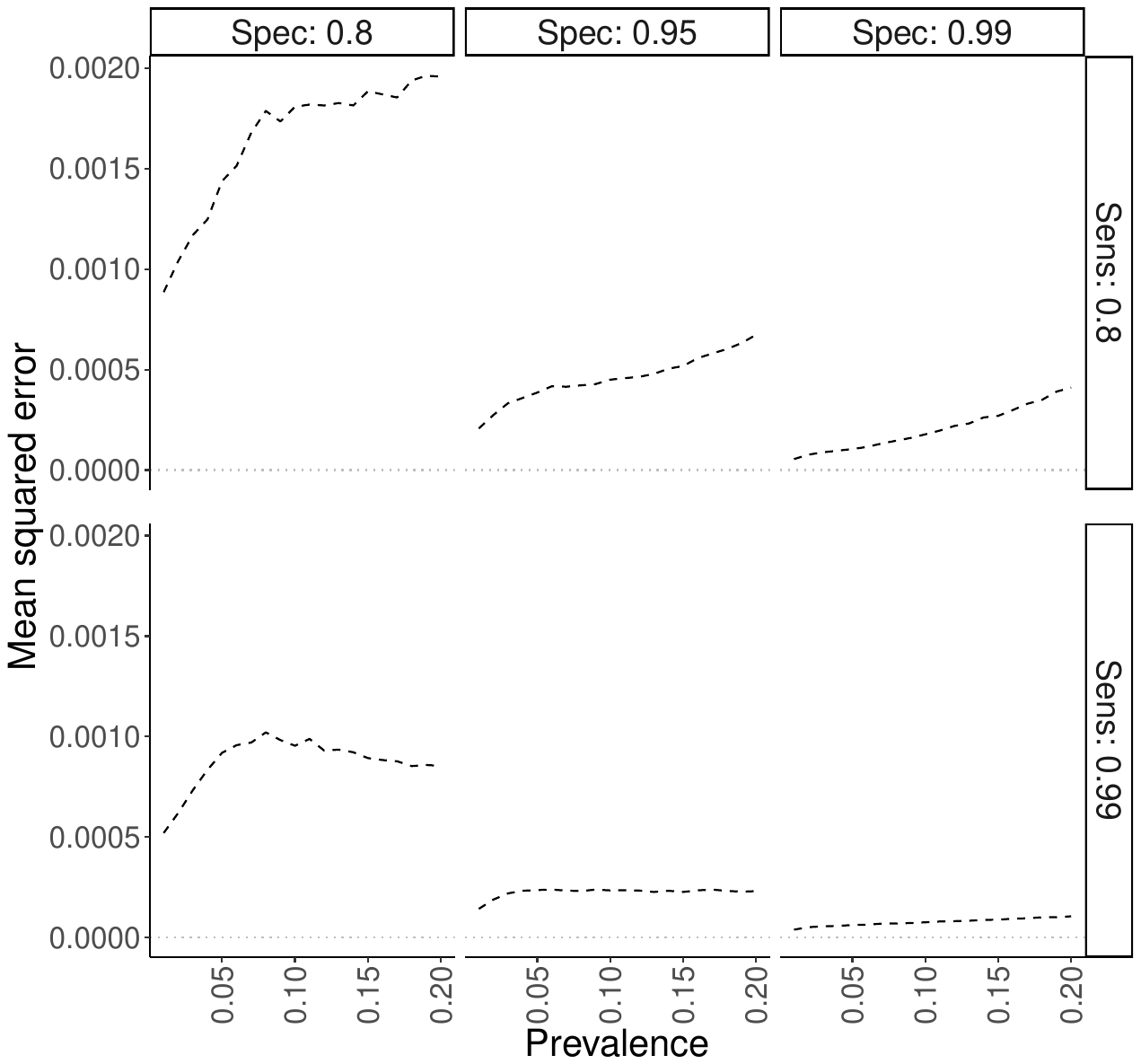}
				\caption{Mean squared error of the Rogan-Gladen ($\hat \pi_{RG}$) estimator from simulation study for DGP 1, described in Section 4.1 of the main text. 10,000 simulations were conducted for this scenario.}
				\label{fig:dgp1_mse}
			\end{figure}
			
			\begin{figure}
				\centering
				\includegraphics[width=1\textwidth]{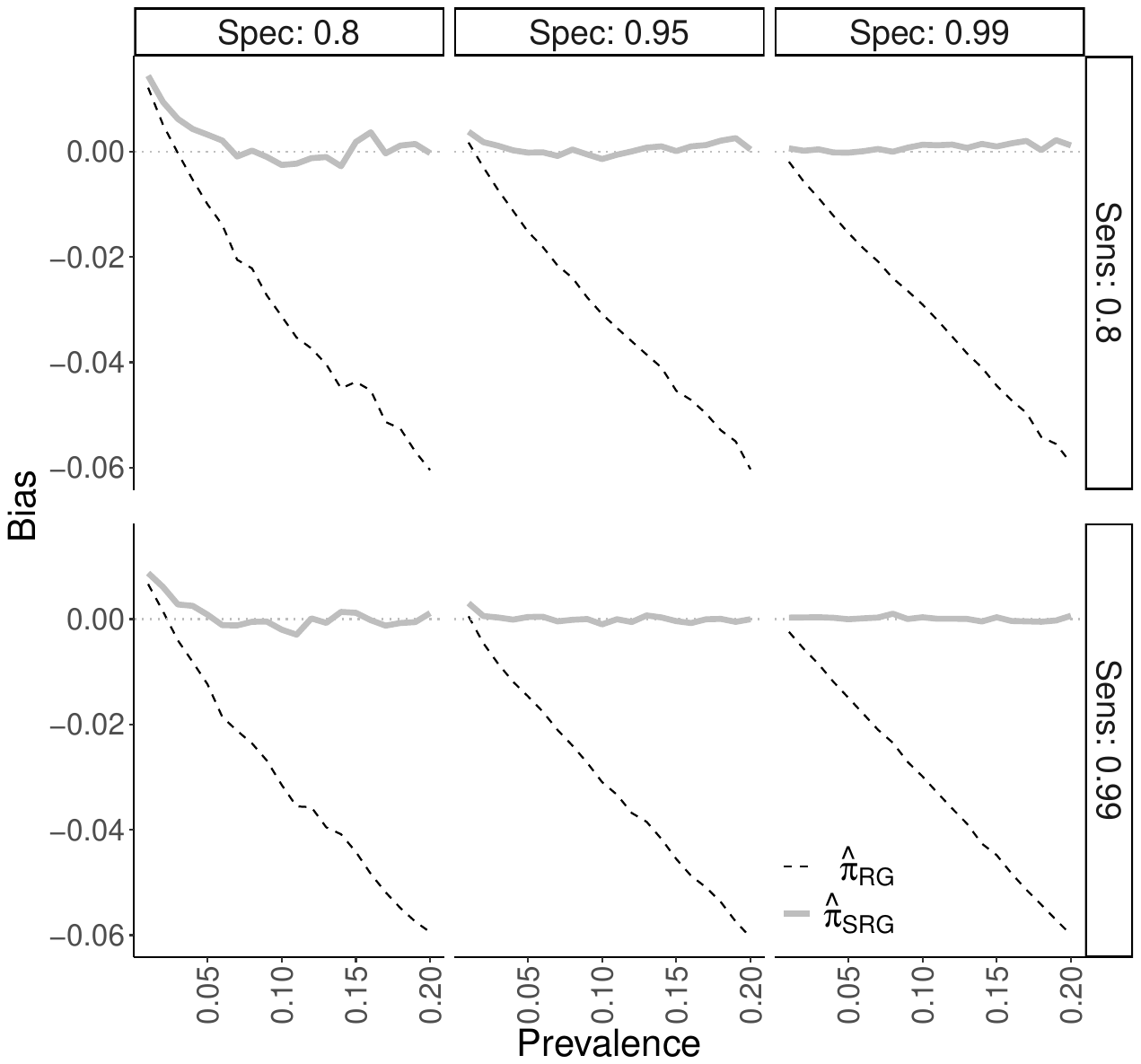}
				\caption{	Empirical bias of the Rogan-Gladen ($\hat \pi_{RG}$) and nonparametric standardized ($\hat \pi_{SRG}$) estimators from simulation study for DGP 2, described in Section 4.2 of the main text.}
				\label{fig:dgp2_bias}
			\end{figure}
			
			\begin{figure}
				\centering
				\includegraphics[width=1\textwidth]{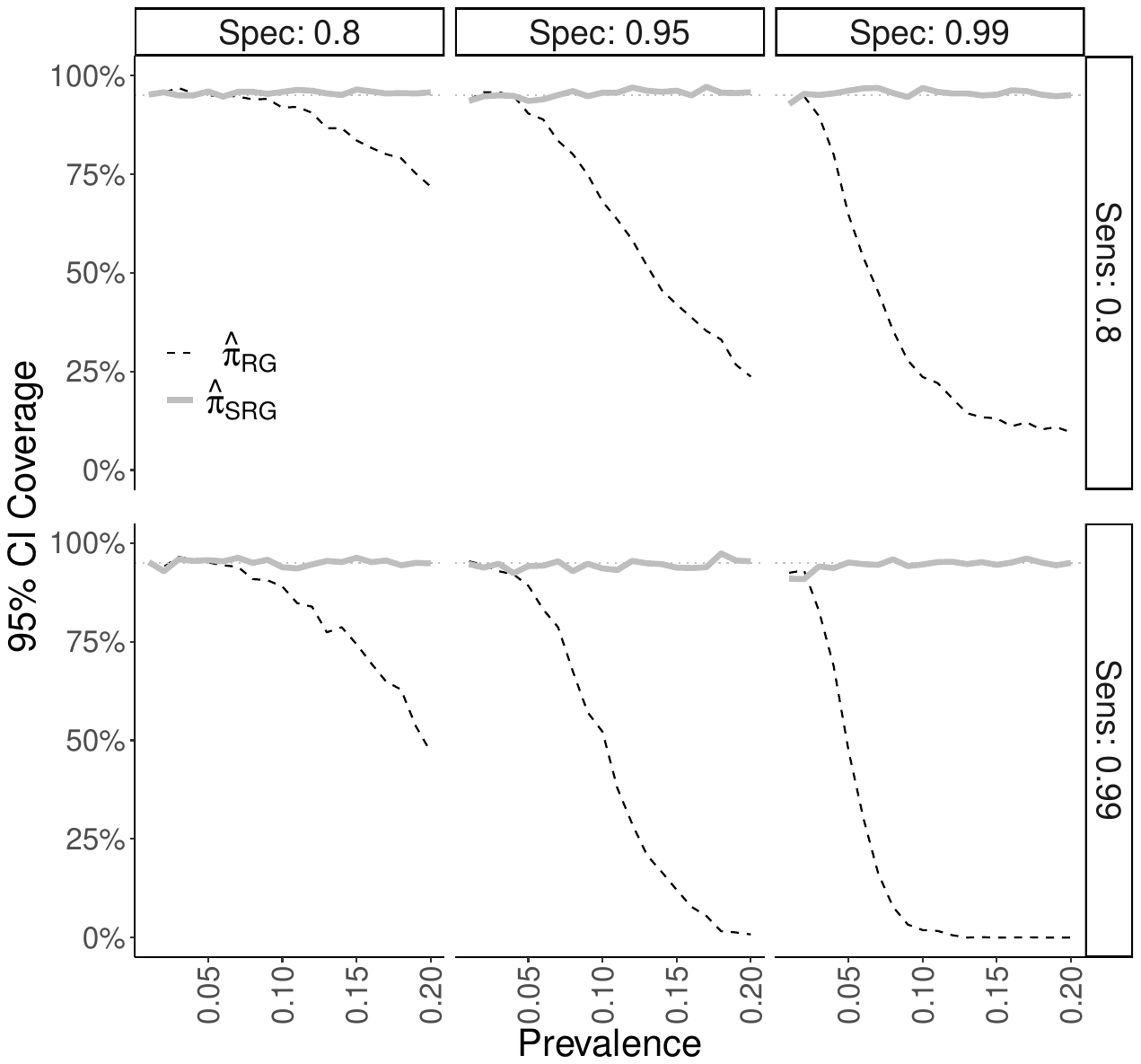}
				\caption{Confidence interval coverage of the Rogan-Gladen ($\hat \pi_{RG}$) and nonparametric standardized ($\hat \pi_{SRG}$) estimators from simulation study for DGP 2, described in Section 4.2 of the main text.}
				\label{fig:dgp2_coverage}
			\end{figure}
			
			\begin{figure}
				\centering
				\includegraphics[width=1\textwidth]{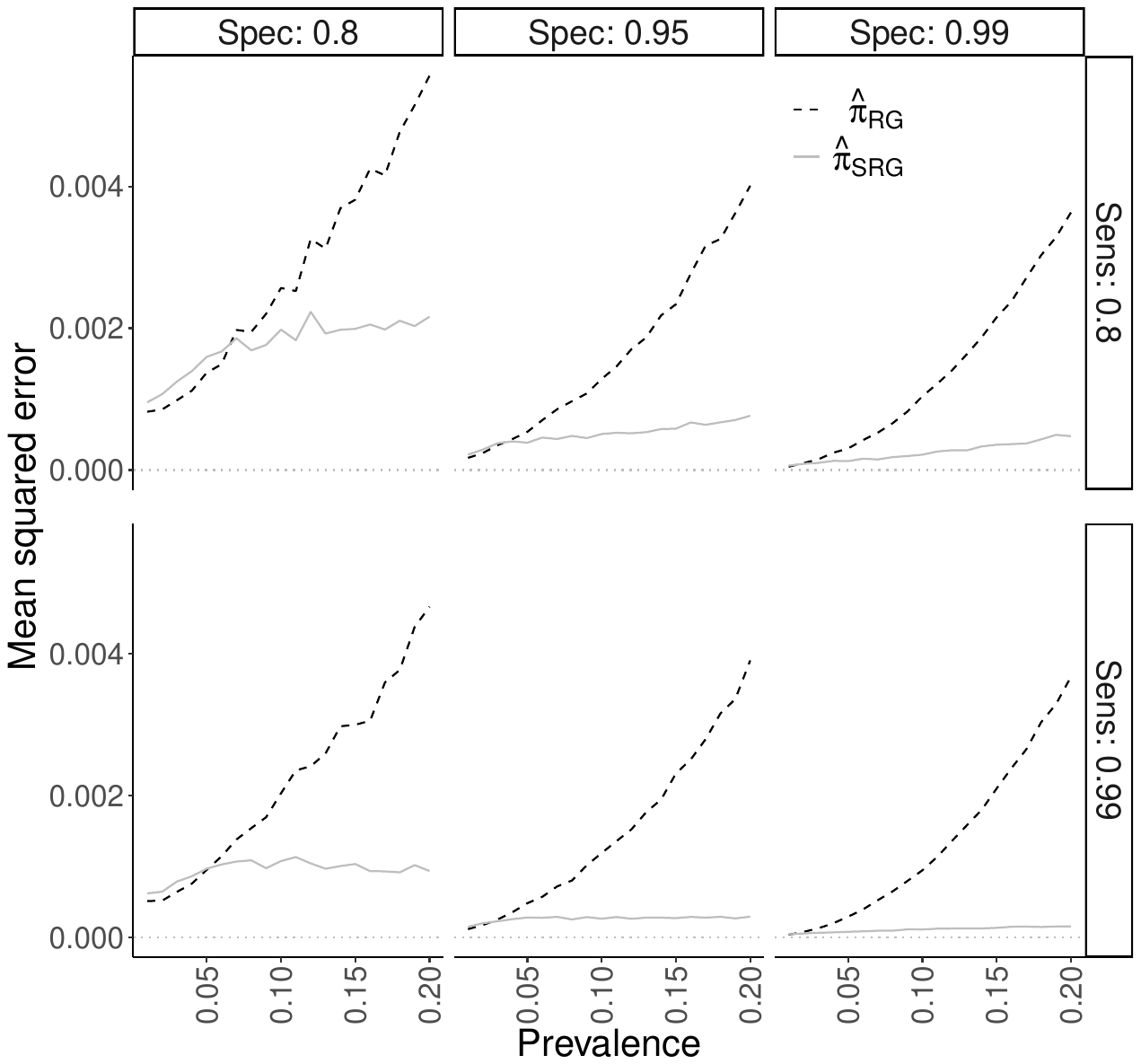}
				\caption{Mean squared error of the Rogan-Gladen ($\hat \pi_{RG}$) and nonparametric standardized ($\hat \pi_{SRG}$) estimators from simulation study for DGP 2, described in Section 4.2 of the main text.}
				\label{fig:dgp2_mse}
			\end{figure}
			
			\begin{figure}
				\centering
				\includegraphics[width=1\textwidth]{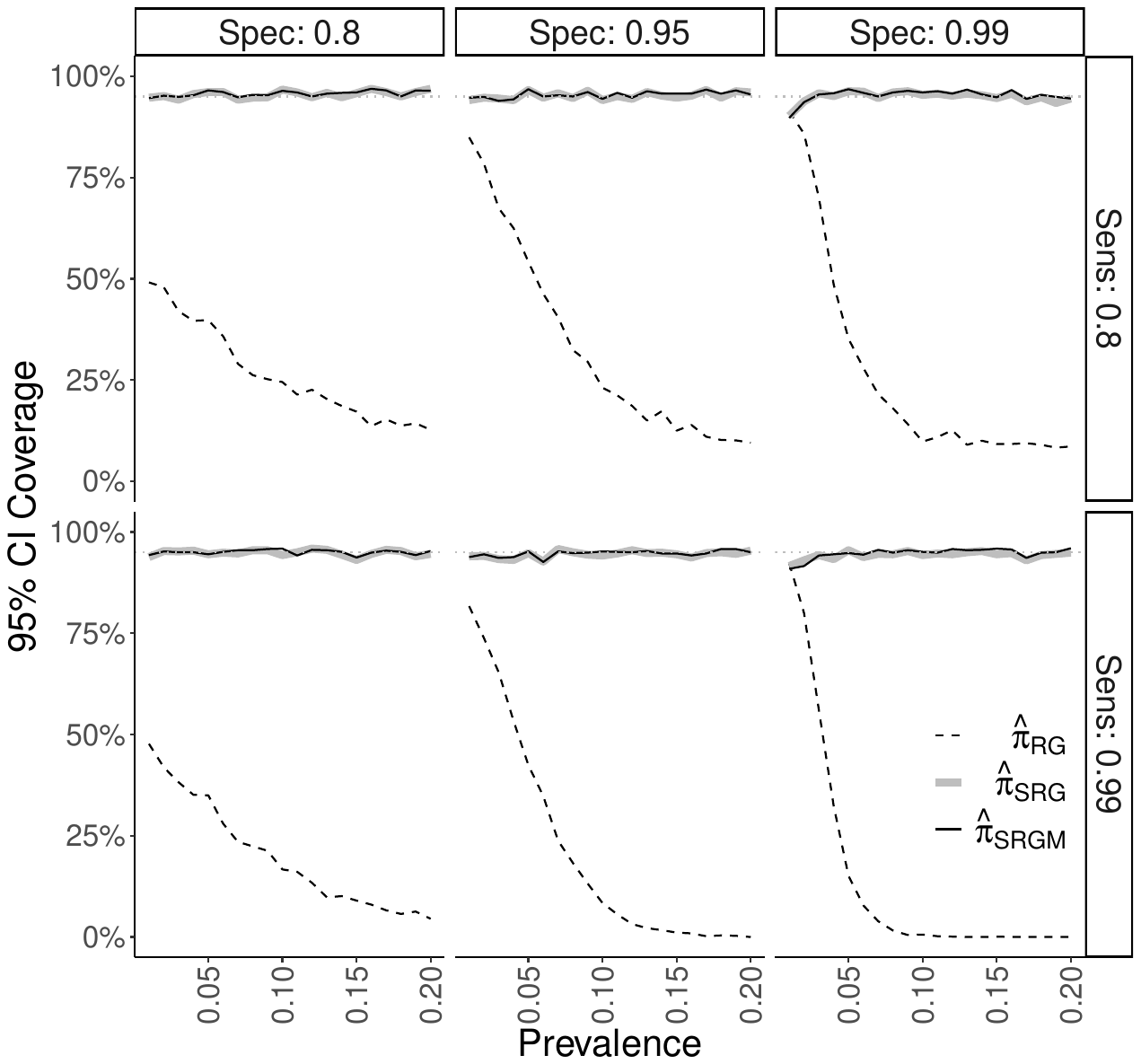}
				\caption{Confidence interval coverage of the Rogan-Gladen ($\hat \pi_{RG}$), nonparametric standardized ($\hat \pi_{SRG}$), and parametric standardized ($\hat \pi_{SRGM}$) estimators from simulation study for DGP 3, described in Section 4.3.1 of the main text.}
				\label{fig:dgp3_coverage}
			\end{figure}
			
			\clearpage

			\begin{figure}
				\centering
				\includegraphics[width=1\textwidth]{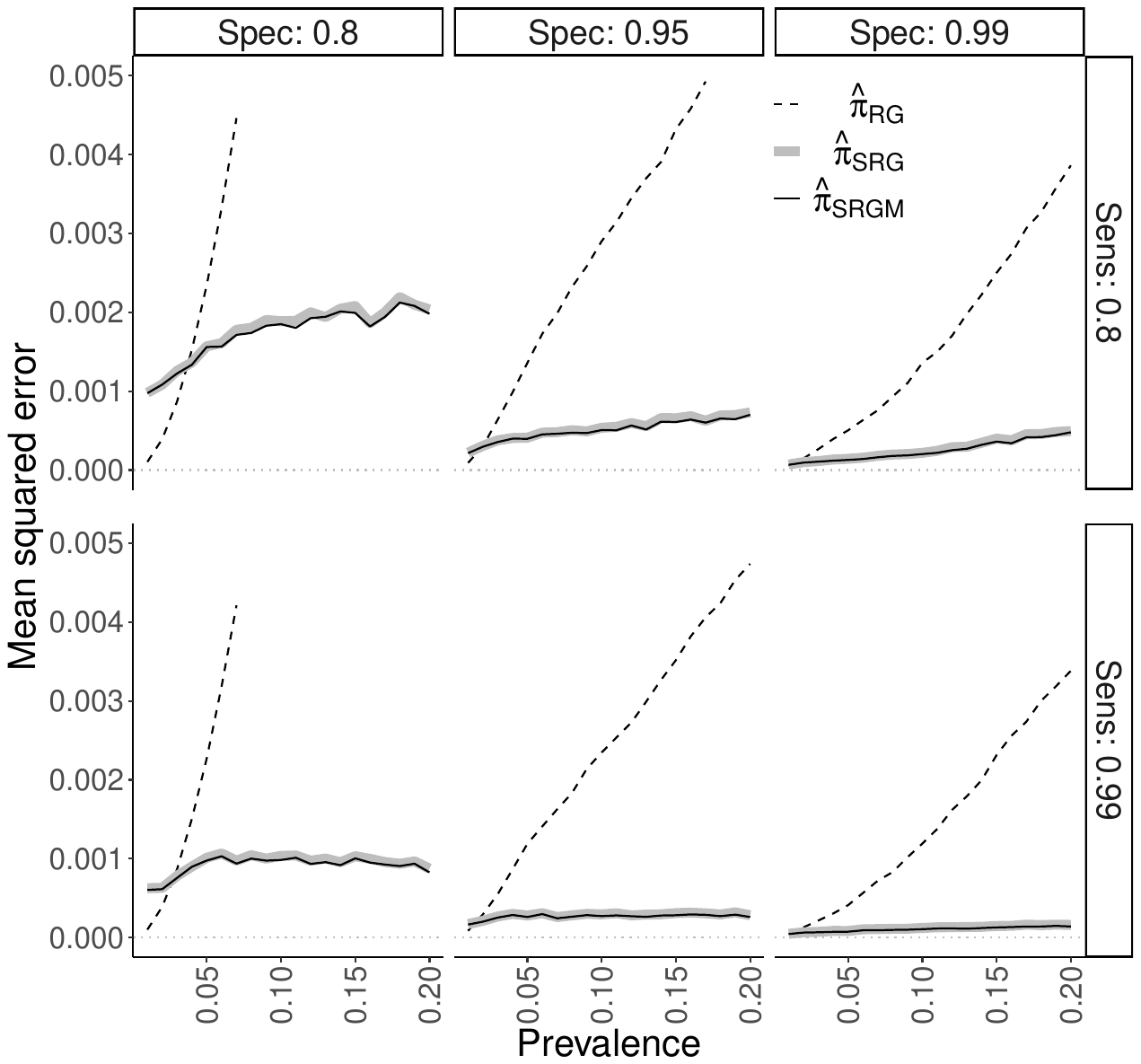}
				\caption{Mean squared error of the Rogan-Gladen ($\hat \pi_{RG}$), nonparametric standardized ($\hat \pi_{SRG}$), and parametric standardized ($\hat \pi_{SRGM}$) estimators from simulation study for DGP 3, described in Section 4.3.1 of the main text. The $y$-axis is truncated at 0.005 for ease of distinguishing $\hat \pi_{SRG}$ and $\hat \pi_{SRGM}$.}
				\label{fig:dgp3_mse}
			\end{figure}
			
			\clearpage
			
			\begin{figure}
				\centering
				\includegraphics[width=1\textwidth]{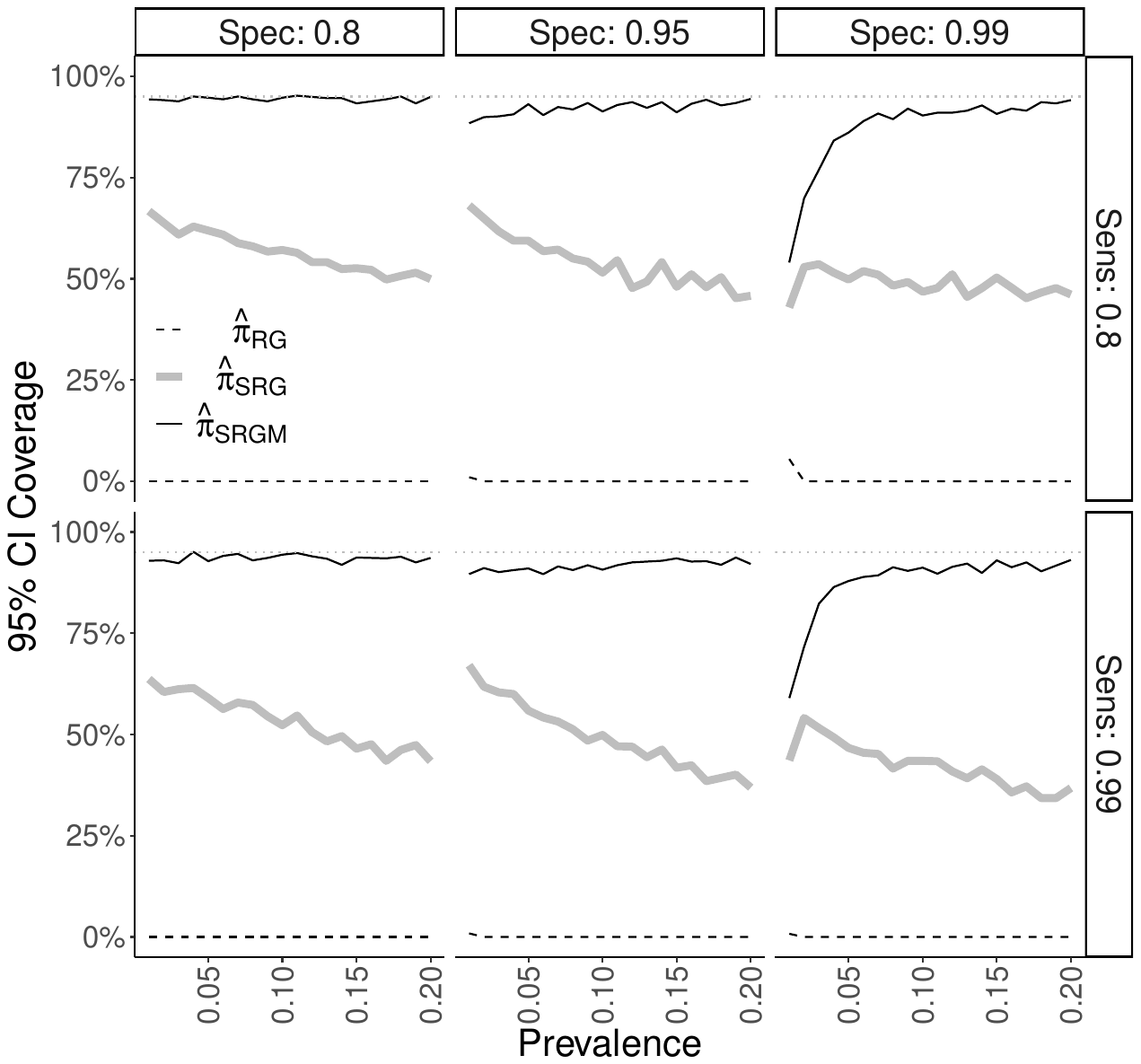}
				\caption{Confidence interval coverage of the Rogan-Gladen ($\hat \pi_{RG}$), nonparametric standardized ($\hat \pi_{SRG}$), and parametric standardized ($\hat \pi_{SRGM}$) estimators from simulation study for DGP 4, described in Section 4.3.2 of the main text.}
				\label{fig:dgp4_coverage}
			\end{figure}
			
			\clearpage

			\begin{figure}
				\centering
				\includegraphics[width=1\textwidth]{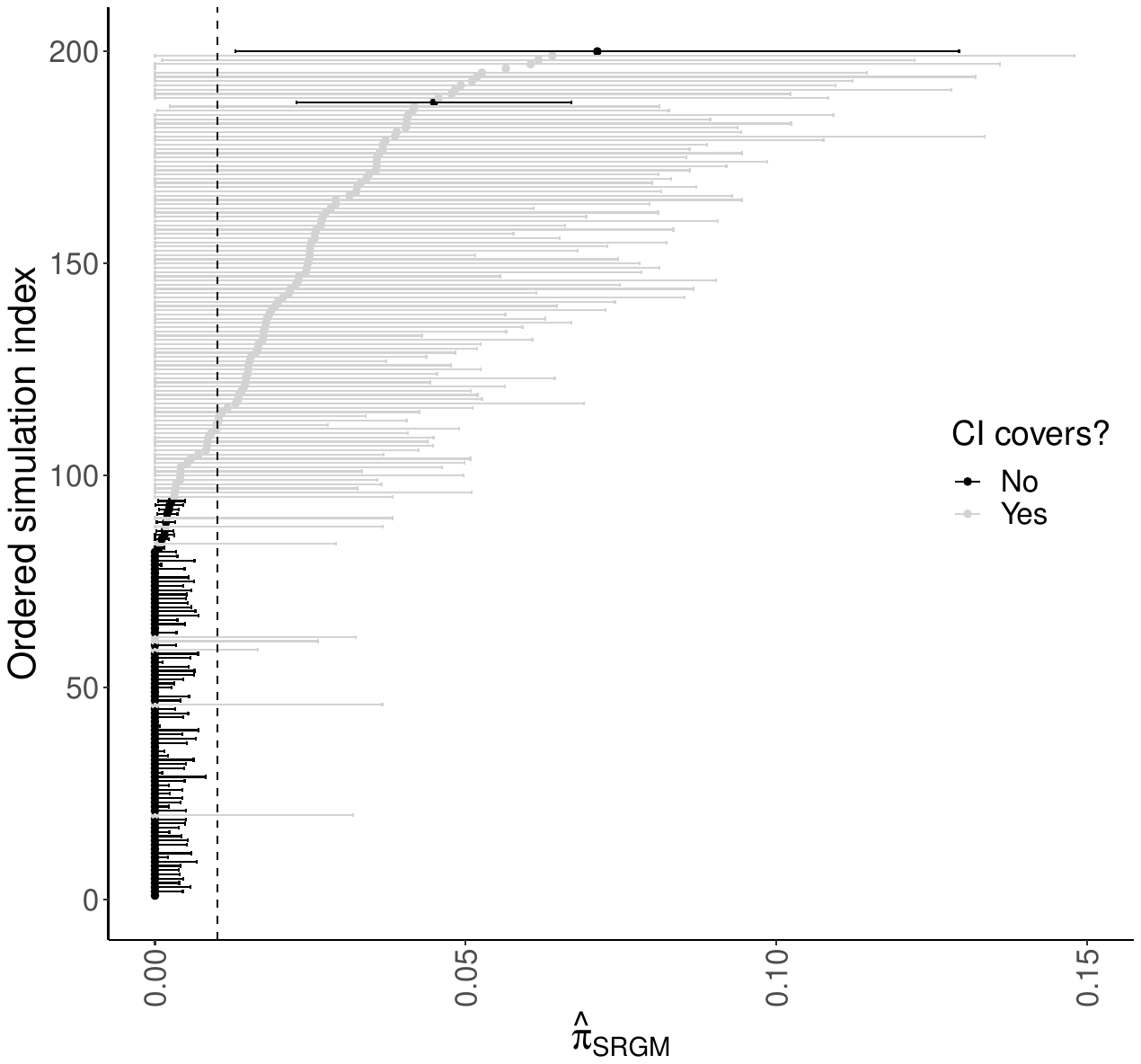}
				\caption{A random sample of 200 point estimates $\hat \pi_{SRGM}$ and 95\% confidence interval estimates based on $\hat V_{\pi, SRGM}$ from DGP 4, where the data were generated with $\sigma_e=\sigma_p=.99$ and $\pi = .01$. 59\% of the intervals covered the true value of $\pi=.01$. The $x$-axis is truncated at 0.015 for visibility of the estimates.}
				\label{fig:dgp4_diagnostics}
			\end{figure}
			
			\clearpage

			\begin{figure}
				\centering
				\includegraphics[width=1\textwidth]{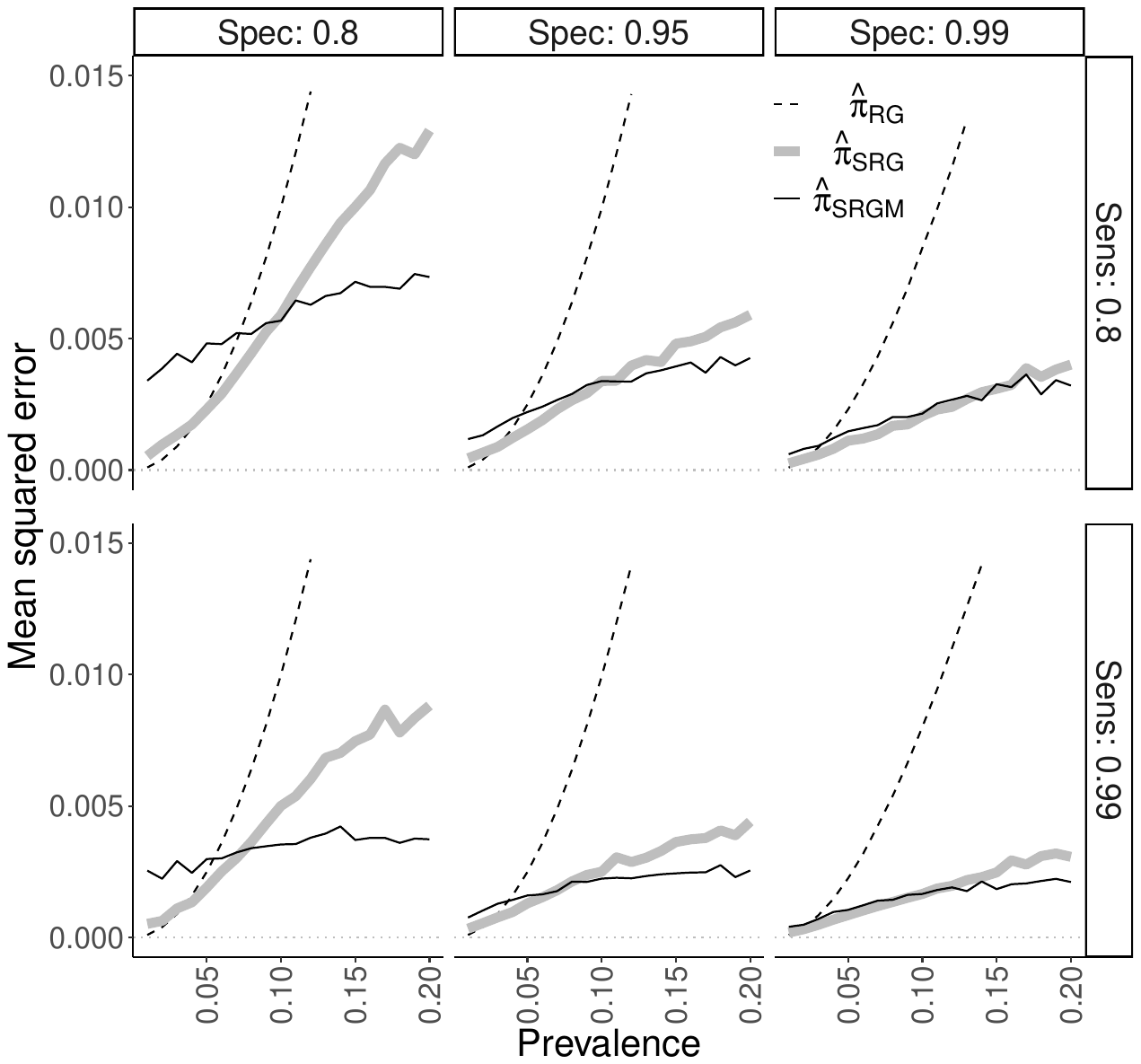}
				\caption{Mean squared error of the Rogan-Gladen ($\hat \pi_{RG}$), nonparametric standardized ($\hat \pi_{SRG}$), and parametric standardized ($\hat \pi_{SRGM}$) estimators from simulation study for DGP 4, described in Section 4.3.1 of the main text. The $y$-axis is truncated at 0.015 for ease of distinguishing $\hat \pi_{SRG}$ and $\hat \pi_{SRGM}$.}
				\label{fig:dgp4_mse}
			\end{figure}
			
			\clearpage
			
			\begin{figure}
				\centering
				\includegraphics[width=1\textwidth]{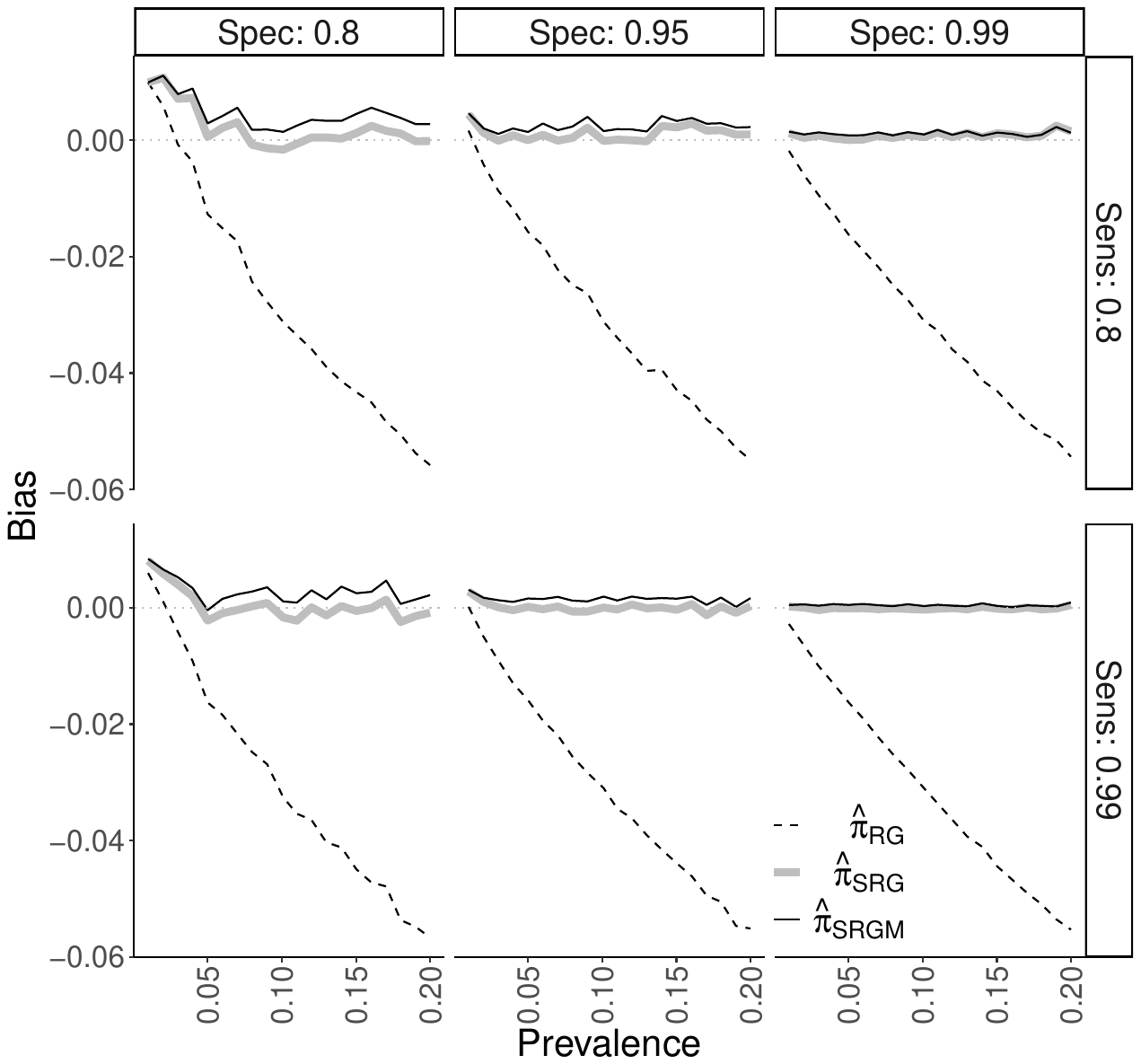}
				\caption{Empirical bias of the estimators from simulation study for DGP 3 under model misspecification, described in Section 4.4 of the main text.}
				\label{fig:dgp5}
			\end{figure}
			
			\clearpage 
			
			\begin{figure}
				\centering
				\includegraphics[width=1\textwidth]{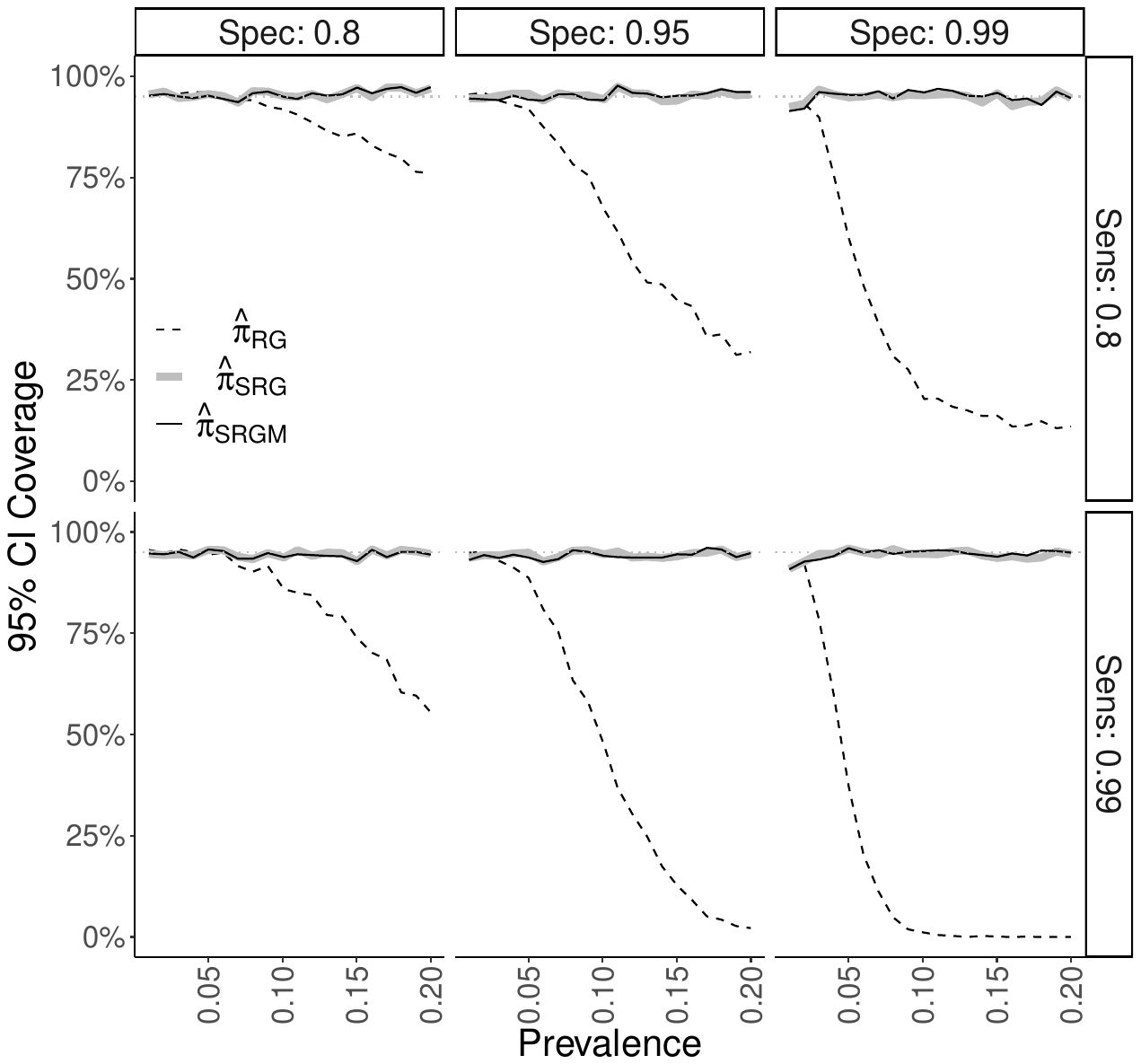}
				\caption{Confidence interval coverage of the estimators from simulation study for DGP 3 under model misspecification, described in Section 4.4 of the main text.}
				\label{fig:dgp5_coverage}
			\end{figure}
			
			\clearpage 
			
			\begin{figure}
				\centering
				\includegraphics[width=1\textwidth]{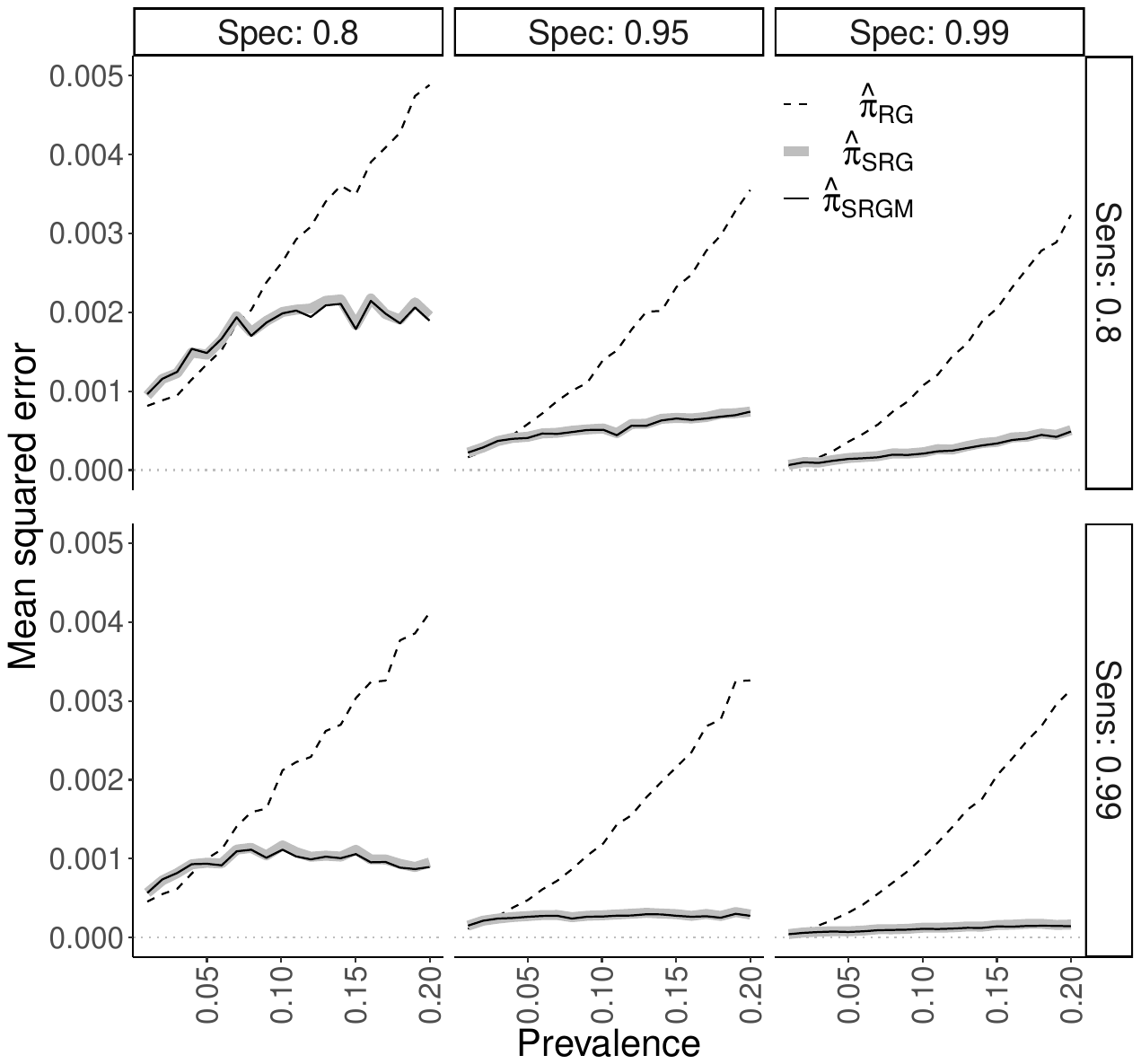}
				\caption{Mean squared error of the Rogan-Gladen ($\hat \pi_{RG}$), nonparametric standardized ($\hat \pi_{SRG}$), and parametric standardized ($\hat \pi_{SRGM}$) estimators from simulation study for DGP 3 under model misspecification, described in Section 4.4 of the main text. The $y$-axis is truncated at 0.005 for ease of distinguishing $\hat \pi_{SRG}$ and $\hat \pi_{SRGM}$.}
				\label{fig:dgp5_mse}
			\end{figure}
			
			\begin{figure}
				\centering
				\includegraphics[width=1\textwidth]{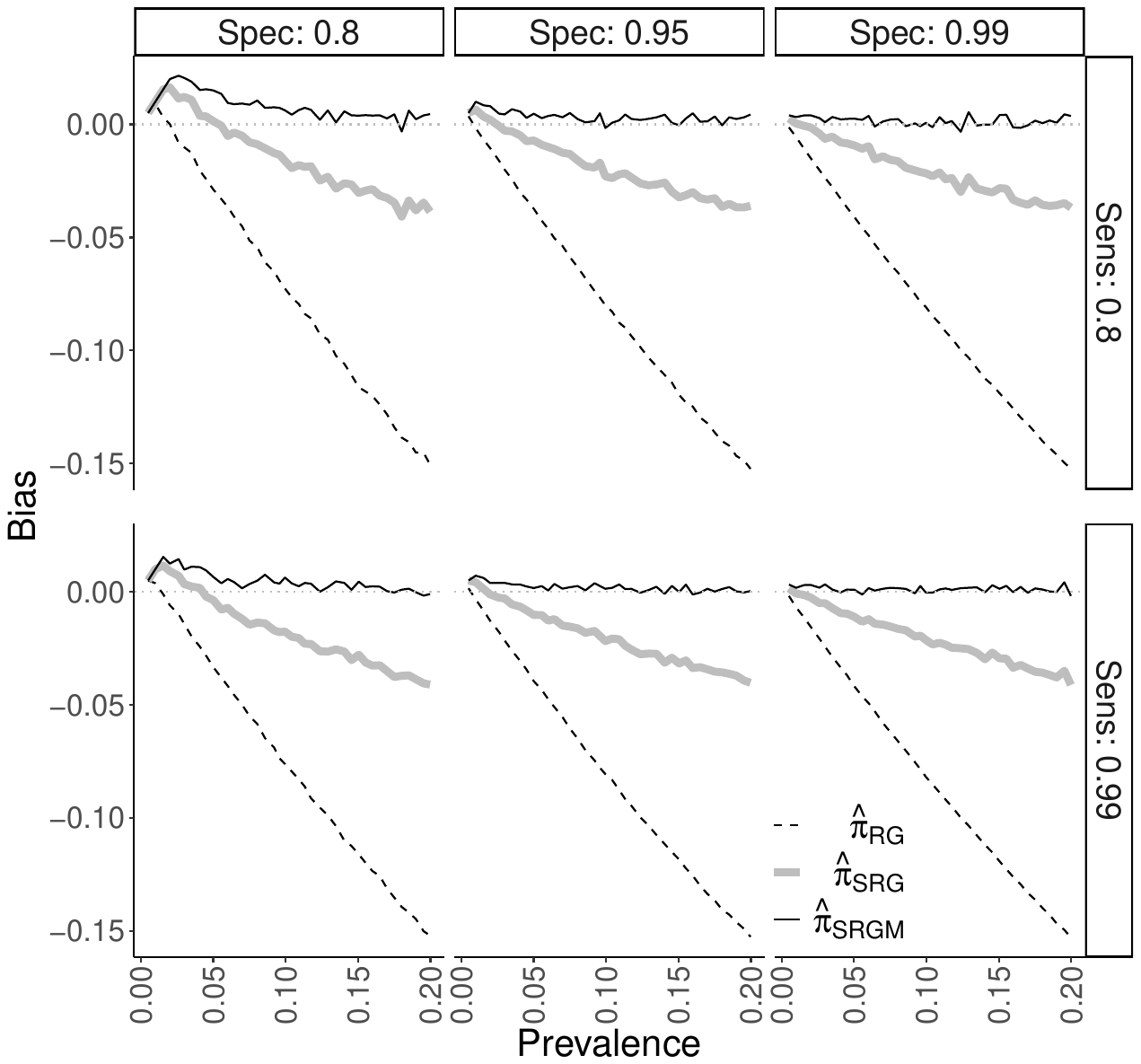}
				\caption{Empirical bias of the estimators from simulation study for DGP 4 under model misspecification, described in Section 4.4 of the main text.}
				\label{fig:dgp6}
			\end{figure}
			
			\clearpage 
			
			\begin{figure}
				\centering
				\includegraphics[width=1\textwidth]{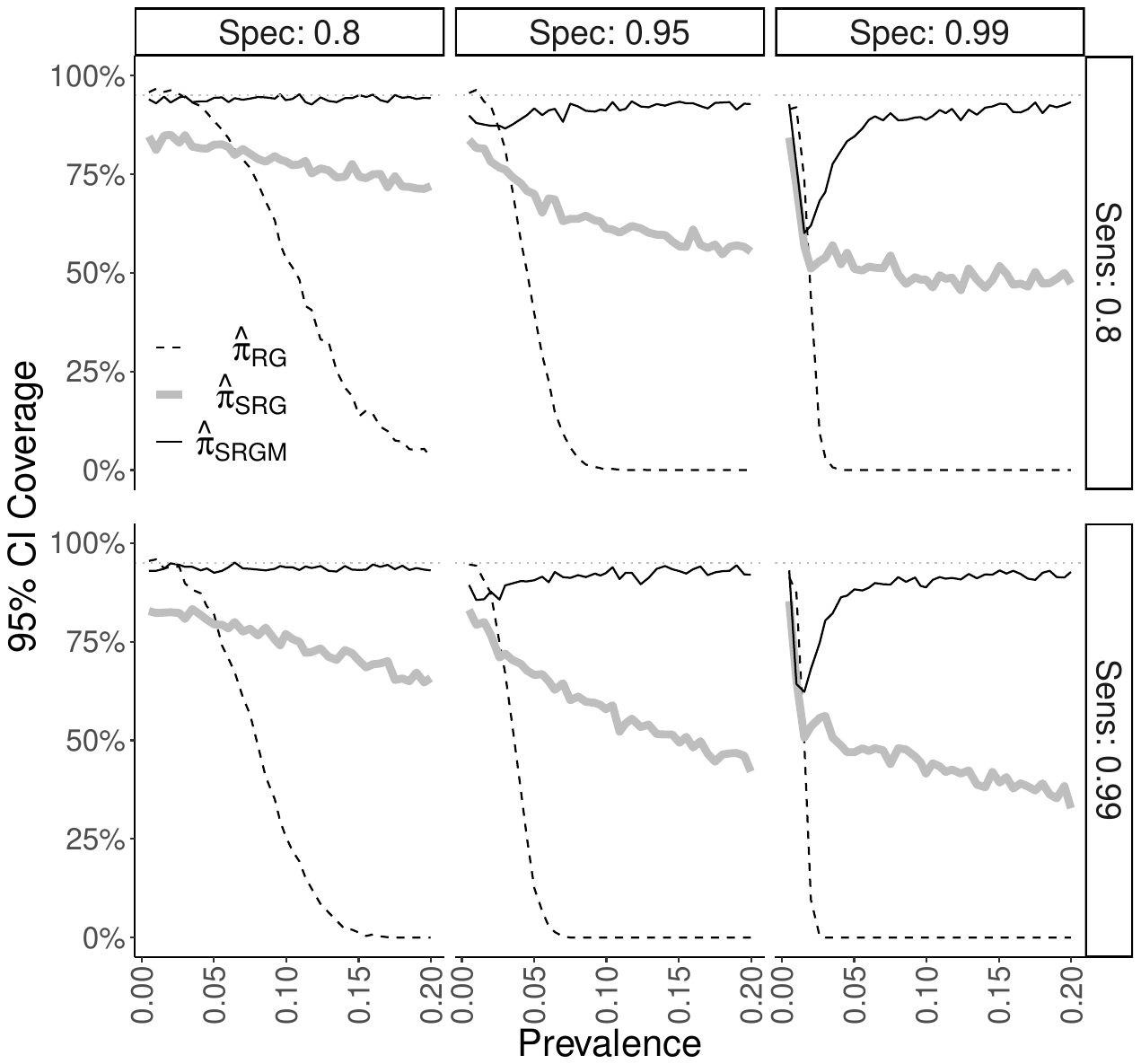}
				\caption{Confidence interval coverage of the estimators from simulation study for DGP 4 under model misspecification, described in Section 4.4 of the main text.}
				\label{fig:dgp6_coverage}
			\end{figure}
			
			\begin{figure}
				\centering
				\includegraphics[width=1\textwidth]{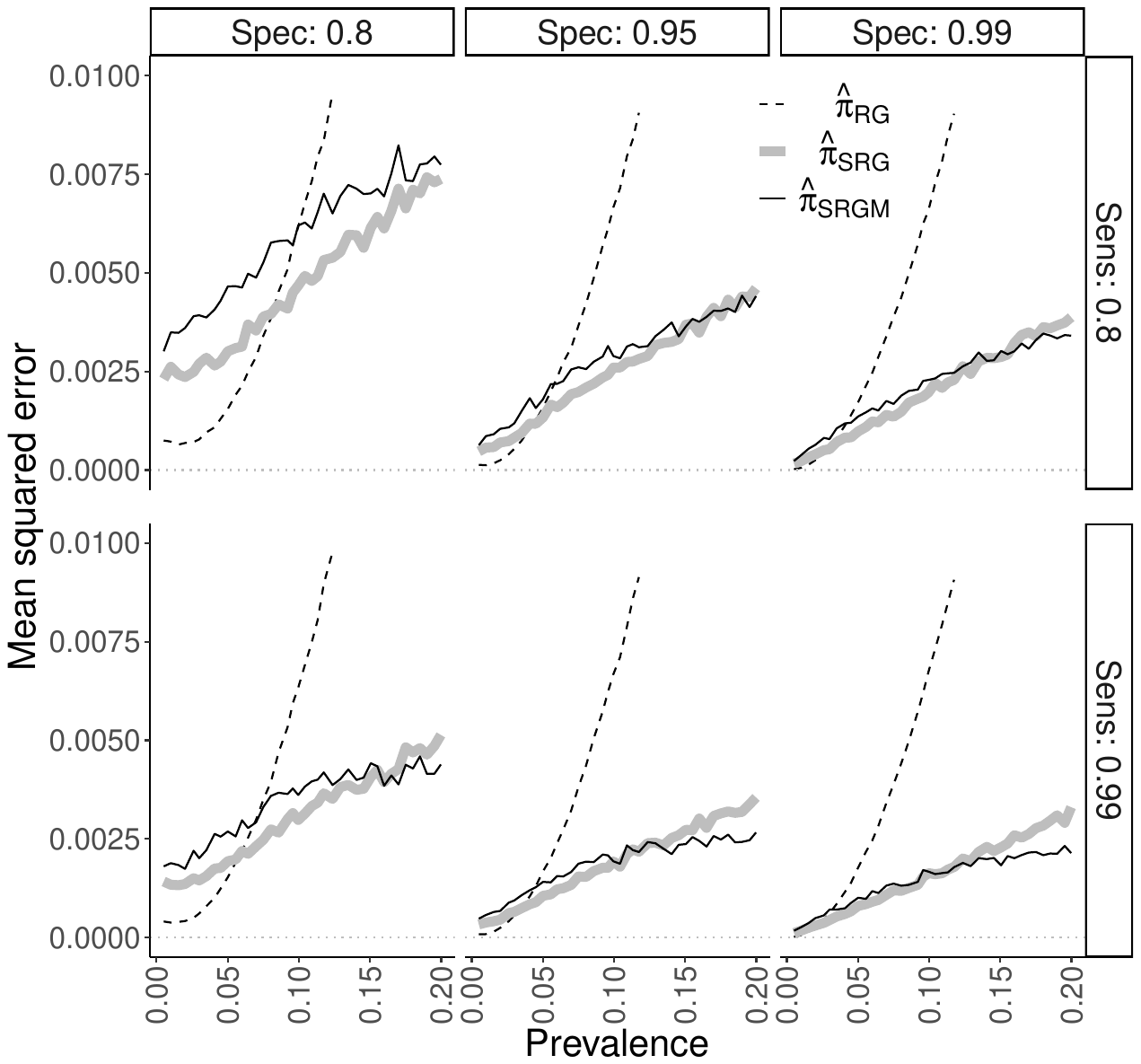}
				\caption{Mean squared error of the Rogan-Gladen ($\hat \pi_{RG}$), nonparametric standardized ($\hat \pi_{SRG}$), and parametric standardized ($\hat \pi_{SRGM}$) estimators from simulation study for DGP 4 under model misspecification, described in Section 4.4 of the main text. The $y$-axis is truncated at 0.01 for ease of distinguishing $\hat \pi_{SRG}$ and $\hat \pi_{SRGM}$.}
				\label{fig:dgp6_mse}
			\end{figure}
			
			\begin{figure}
				\centering
				\includegraphics[width=1\textwidth]{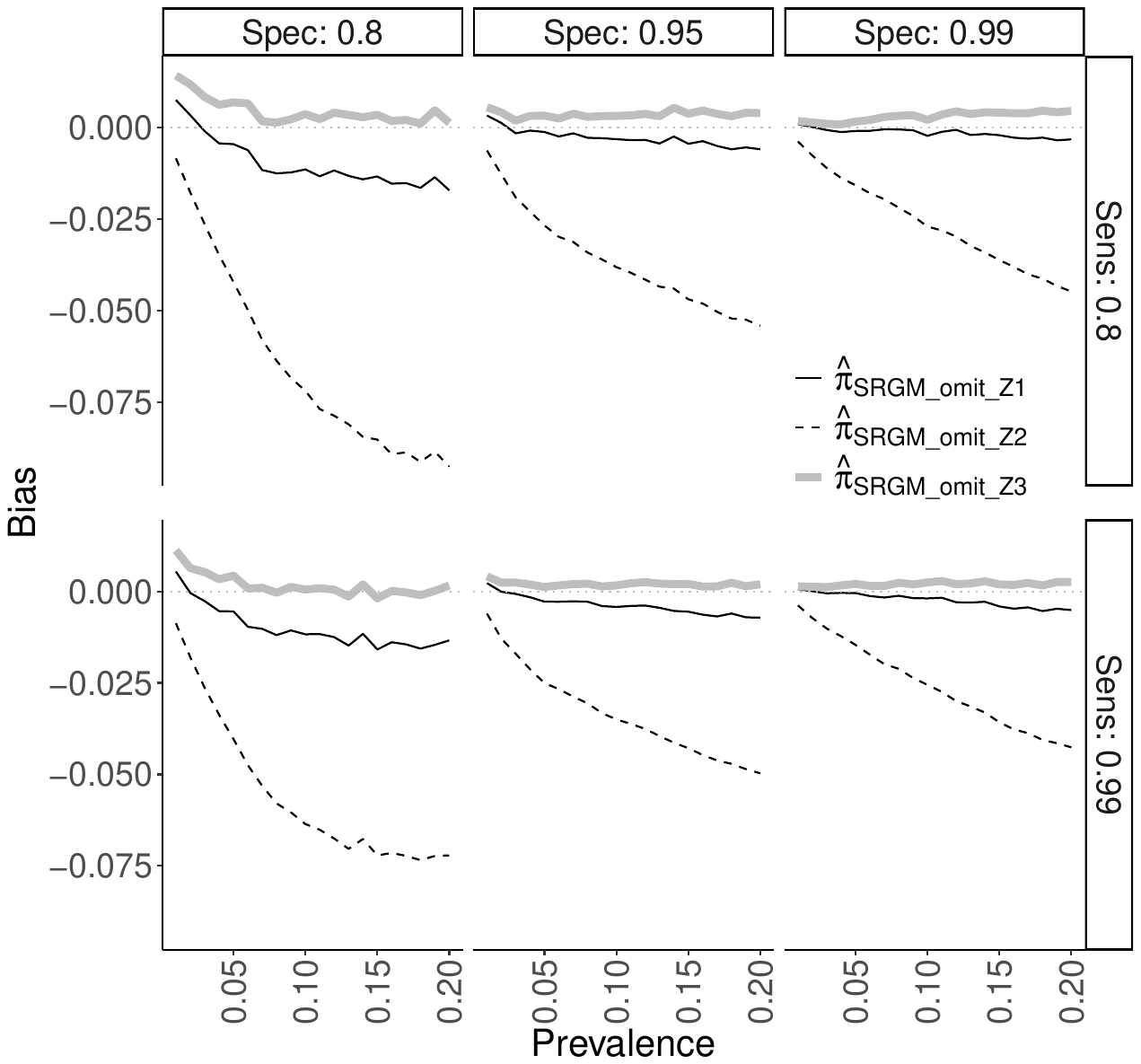}
				\caption{Empirical bias in simulation study for DGP 3 with the model-based estimator $\hat \pi_{SRGM}$ misspecified by omitting $Z_1$, $Z_2$, and $Z_3$, respectively, described in Section 4.4 of the main text.}
				\label{fig:dgp10}
			\end{figure}\vspace{.4cm}
				
				
				\section*{Appendix References}
					Boos, D. D., \& Stefanski, L. A. (2013). \textit{Essential Statistical Inference}. Springer New York, Ch. 7, pp. 297-337. 
					
					Miller, K. S. (1981). On the inverse of the sum of matrices. \textit{Mathematics Magazine}, 54(2), 67-72.
					
			\end{appendices}
			
		\end{document}